\documentclass[aps,pre,floatfix,footinbib]{revtex4}

\usepackage{natbib}
\usepackage{xcolor}
\usepackage{hyperref}        % hyperlinks
\usepackage{url}            % simple URL typesetting
\usepackage{booktabs}       % professional-quality tables
\usepackage{amsfonts}       % blackboard math symbols
\usepackage{nicefrac}       % compact symbols for 1/2, etc.
\usepackage{microtype}      % microtypography
\usepackage{lipsum}
\usepackage{graphicx}
\graphicspath{ {./images/} }
\usepackage{float}
\usepackage{amsmath}
\newcommand{\be}{\begin{equation}}
\newcommand{\ee}{\end{equation}}
\newcommand{\eq}[1]{(\ref{#1})}

\begin{document}

%\title{Non-backtracking walks reveal functional communities in sparse structural connectome of a worm}

\title{Communities in C.elegans connectome through the prism of non-backtracking walks}

\author{Arsenii A. Onuchin$^{1,2}$, Alina V. Chernizova$^3$, Mikhail A. Lebedev$^4, ^5$, Kirill E. Polovnikov$^{1,\dagger}$}

\affiliation{$^1$Skolkovo Institute of Science and Technology, 121205 Moscow, Russia \\
$^2$Laboratory of Complex Networks, Center for Neurophysics and Neuromorphic Technologies, Moscow, Russia \\
$^3$Institute of Higher Nervous Activity
and Neurophysiology of the Russian Academy of Sciences, 117485 Moscow, Russia \\
$^4$Faculty of Mechanics and Mathematics, Lomonosov Moscow State University, 119991 Moscow, Russia\\
$^5$Sechenov Institute of Evolutionary Physiology and Biochemistry of the Russian Academy of Sciences, 194223 Saint Petersburg, Russia\\
$^\dagger$Corresponding author: kipolovnikov@gmail.com}

\begin{abstract}
\section*{Abstract}
The fundamental relationship between the mesoscopic structure of neuronal circuits and organismic functions they subserve is one of the major challenges in contemporary neuroscience. Formation of structurally connected modules of neurons enacts the conversion from single-cell firing to large-scale behaviour of an organism, highlighting the importance of their accurate profiling in the data. While connectomes are typically characterized by significant sparsity of neuronal connections, recent advances in network theory and machine learning have revealed fundamental limitations of traditionally used community detection approaches in cases where the network is sparse. Here we studied the optimal community structure in the structural connectome of \textit{C.elegans}, for which we exploited a non-conventional approach that is based on non-backtracking random walks, virtually eliminating the sparsity issue. In full agreement with the previous asymptotic results, we demonstrated that non-backtracking walks resolve the ground truth annotation into clusters on stochastic block models (SBM) with the size and density of the connectome better than the spectral methods related to simple random walks. Based on the cluster detectability threshold, we determined that the optimal number of modules in a recently mapped connectome of \textit{C.elegans} is 10, which precisely corresponds to the number of isolated eigenvalues in the spectrum of the non-backtracking flow matrix. The discovered communities have a clear interpretation in terms of their functional role, which allows one to discern three structural compartments in the worm: the Worm Brain (WB), the Worm Movement Controller (WMC), and the Worm Information Flow Connector (WIFC). Broadly, our work provides a robust network-based framework to reveal mesoscopic structures in sparse connectomic datasets, paving way to further investigation of connectome mechanisms for different functions.  %We expect that broad application of the non-backtracking walks to sparse connectomes of other organisms (especially, human) would foster understanding of the modular organisation as the basis for collective neuronal behaviour.

%Spectral clustering is one of the most popular methods for community detection in biological networks. The problem of dense graph clustering is quite well solved by standard spectral methods, while sparse graph clustering has significant difficulties. Here, we estimate a sparse graph community structure with a non-backtracking clustering approach, applying it to the \textit{Caenorhabditis elegans} (\textit{C. elegans}) nervous system (connectome) as a model system, the well-known real-world sparse graph. We have shown that the mentioned method surpasses other clustering methods in the detection of SBM clusters and successfully finds communities in the C.elegans connectome. The number of clusters was set taking clustering resolution limit into account, with new limit-resolution method described. The resulting communities have good biological interpretability in terms of ganglia and functional neuronal groups, which allows us to speak about the existence of an explicit mathematical logic and organisation principle lying behind the organization of the worm nervous system. We expect that a new approach may be useful for investigating the sparse complex network community structure in other (nervous) systems.

\end{abstract}

\maketitle

% keywords can be removed

\section*{Introduction}

Complexity of biological and social systems driven by collective behaviour of their agents is commonly studied using network (or graph) representation, where nodes represent agents and edges correspond to pairwise coupling between them. The resulting dimensionality reduction frequently allows to extract the most valuable information about hidden relationships governing static and dynamic properties of a system. One of the most striking and practically important examples of such information is the mesoscopic organization of the network in modules or communities.

The nervous system is no exception in this regard as it can be represented as a structural connectome, that is, a graph, where vertices are nerve cells and edges reflect direct structural connections (wiring) between them. Similarly to most of real-world networks, the connectome is extremely sparse, that is, its number of theoretically possible connections between neurons greatly exceeds the factual amount of connections \cite{karrer2014percolation}.

Such a reduction of excessive edges is a consequence of network \textit{modularity}, a tendency to form assortative communities (modules) with relatively loose inter-connections. Like an effective team work of people where complex problem requires distribution of tasks among specialized groups of participants, mesoscopic organization of neurons in a connectome serves to facilitate certain functions of the nervous system, such as "fire together wire together" principle, \cite{hebb1949first}. Thus, accurate detecting of modules (communities) in the connectome data can help to establish a conversion between micro-level single neuron interactions and macro-level organism behaviour.

Community detection is an extremely hot topic in various fields such as technological \cite{albert99,broder00}, biological \cite{dekker13,ravasz02}, social \cite{redner98,chen09}
and economical \cite{piccardi10,polovnikov_jphysa20,polovnikov_ijcai22} fields. A widely used approach in community detection is a spectral decomposition of a linear
operator defined on a network: information about communities is then encoded in several leading eigenvectors \cite{Von2007tutorial,krivelevich03}. It was shown that all commonly used matrices (adjacency, Laplacian, modularity, non-backtracking, see Methods) readily classify nodes as long as the network density is sufficient \cite{nadakuditi12,decelle11}.
In particular, the modularity operator is one of the most efficient instruments that successfully detects communities in stochastic networks of various nature \cite{newman06,chen09,norton18,salespardo07}. The modularity operator can be used to extract mesoscopic organization in \textit{C.elegans} \cite{pan10}.

Sparse graphs are a special case where most of the traditional community detection methods suffer from fundamental limitations. Namely, at a given cluster strength there is a critical network density below which community detection becomes a very difficult problem \cite{Krzakala2013spectral}. Furthermore,
traditionally used operators (adjacency, Laplacian, modularity) turn out to fail above this threshold, since their leading eigenvectors rapidly become uncorrelated with the intrinsic community structure upon decrease of network density. This behaviour is explained as the emergence of vertices with anomalously high degree (hubs), which eventually perturbs the spectral edge of these operators because of the “Lifshitz tails” in the spectral density of sparse graphs (\cite{Reichardt06,Arenas08,nechaev18}). Localization on hubs, and not on the communities, is thus a major drawback for all conventional spectral methods in the sparse regime.

To address this issue, Krzakala et al. proposed \cite{Krzakala2013spectral} to make use of the non-backtracking (Hashimoto) random walks on the associated directed graph. By construction, such walks cannot revisit the same node immediately at the following step and, as a result, they do not localize on hubs. The leading eigenvectors of the non-backtracking operator (the transfer matrix of non-backtracking walks) encode for the community structure of the graph up to the theoretical resolution limit \cite{Krzakala2013spectral}. Due to their intrinsic ability to deal with sparse graphs, the non-backtracking walks have received increasing attention in the analysis of biological datasets. Recently we have shown that this approach allows to annotate compartments in the three-dimensional chromatin organization at the single-cell level \cite{polovnikov20}.

\begin{figure}
    \centering
    \includegraphics[width=\columnwidth]{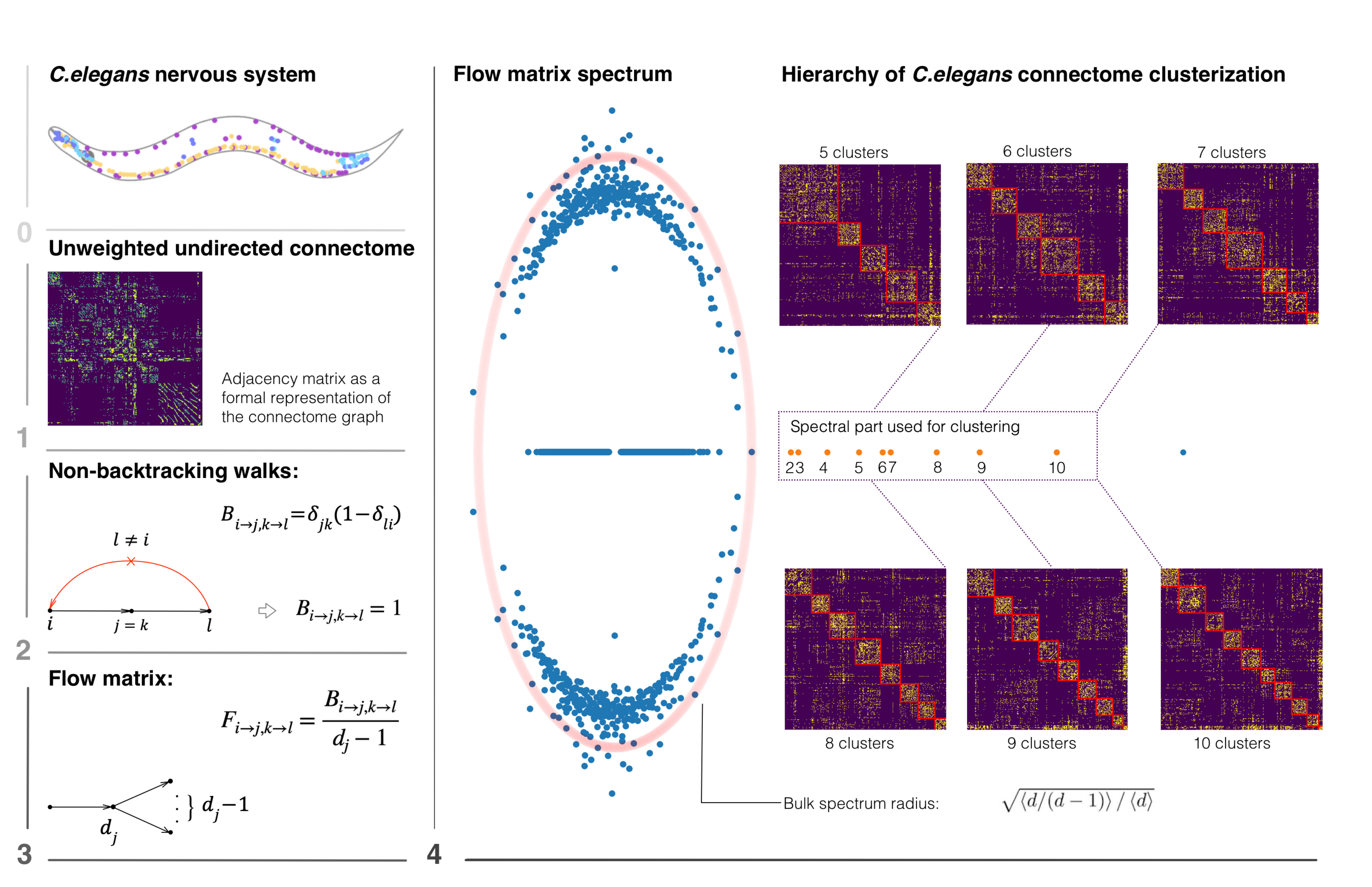}
    \caption{Clusterization of \textit{C.elegans} connectome by means of the spectrum of the non-backtracking flow matrix: \textbf{0}. A simplified scheme of the \textit{C.elegans} nervous system; \textbf{1}. Representation of the connectome as the adjacency matrix; \textbf{2}. Construction of non-backtracking walks on the connectome network; \textbf{3}. Normalization by the out-degree towards the non-backtracking flow operator; \textbf{4}. The spectrum of the flow matrix (new data \cite{cook2019whole}) with the orange dots representing the eigenvalues outside of the spectral radius (pale red), which are used for identification of the communities.}
    \label{fig:pipeline}
\end{figure}

In this work, we examine neuronal connectivity in \textit{Caenorhabditis elegans} — one of the simplest organisms with a structural connectome first mapped by White et al. in 1986 \cite{white1986structure}, which has been completely described by now \cite{cook2019whole}. The nervous system is a prominent part of \textit{C.elegans} with almost one third of all cells in its body being neurons. Importantly, the morphology, location and connectivity of each neuron are remarkably invariant between individuals \cite{hall1991posterior} (it is worth noting that there is now a growing concern on to what extent this is actually true \cite{yemini2021neuropal, brittin2021multi, moyle2021structural, witvliet2021connectomes}), which arguably makes this organism a convenient model for studying neuronal connectivity and related functions. Curiously, due to the elongated shape of the nervous system of the worm, the adjacency matrix of the \textit{C.elegans} connectome is similar to a single-cell chromosome contact map, which describes the spatial proximity of loci in individual conformations of chromosomes \cite{nagano13,polovnikov20}. The so-called scaling in both types of matrices, being a generic polymer (or worm-like chain) feature, is a major source of sparsity in the data.
This prompts one to apply to the \textit{C.elegans} connectome the clustering procedures that has been specifically designed for reliable detection of communities in sparse networks.

Here we study the mesoscopic organization of \textit{C.elegans} connectome by means of the non-backtracking walks. Namely, we construct the Newman's non-backtracking flow operator, which describes the transfer probability of a random walk on the associated directed connectome with prohibited immediate revisiting. The isolated part of the flow matrix spectrum is known to encode for communities and can be used by the clustering algorithm. We ran simulations of community detection on stochastic block models (SBMs) of the corresponding size and density as the connectome and demonstrated that the non-backtracking flow matrix outperformed all traditional operators, in the full agreement with asymptotic results of \cite{Krzakala2013spectral}. In particular, we show better performance of non-backtracking walks over the modularity operator and other approaches, which were previously used for spectral clustering of the \textit{C.elegance} connectome \cite{pan10,pavlovic2014stochastic}.

We consider two \textit{C.elegans} connectome datasets: "old" (Chen et al, 2006) \cite{Chen2006wiring, varshney2011structural} and "new" (Cook et al, 2019) \cite{cook2019whole}. The difference in data completeness between these datasets is quite significant: the number of the edges has approximately doubled in the new connectome (for more details see Methods). We reveal that despite the evident difference in the network density, the modular organization of the two connectomes is rather similar, as reflected by the spectrum of the non-backtracking flow matrix. To establish the detectable amount of communities for each connectome data we propose an algorithm based on the theoretical detectability threshold for SBM-like graphs. In the new data \cite{cook2019whole} our approach reveals $10$ detectable communities in the \textit{C.elegans} connectome, matching the number of isolated eigenvalues in the flow matrix spectrum. The biological interpretation of the communities in the complete connectome suggests that the found clusters highly correlate with the co-localization of neurites (so-called, contactome) in the nervous system of the worm, can be further associated with specific neuronal functions and also overlap with the anatomically defined ganglia. Importantly, we demonstrate that the non-backtracking clusters are much better interpretable than partitions by other spectral algorithms, as well as the modules reported previously \cite{pavlovic2014stochastic,pan2010mesoscopic}. Ultimately our integrative analysis of the mesoscopic structure of the structural connectome and related functions has revealed three neuronal compartments in the \textit{C.elegans}: A. Worm Brain, B. Worm Movements Controller, C. Worm Information Flow Connector.

\section*{Stochastic block model and non-backtracking random walks}

Here we provide a network theory background underlying clustering methods, which is instructive for the definition of the model. The particular connection with several widely used spectral clustering approaches is described in the Methods section. Stochastic block model (SBM) is a commonly used benchmark for community detection in real-world networks, with several important results obtained for asymptotically large networks \cite{decelle2011asymptotic,decelle11,newman06,newman_sparse,newman2006modularity,Krzakala2013spectral}. By definition, a SBM is a generalization of an Erd{\"o}s-Renyi random graph on $N$ nodes, where all edges are generated independently with the probability $p$ that depends on the type of the nodes that it connects. The nodes in a graph belong to $k$ different types (clusters), $G_i, i = 1, 2, ..., k$. Thus, each pair of nodes $(i, j): i \in G_r, j \in G_t$ gets independently connected by an edge with some probability $w_{rt}$, which can be written as a matrix of  pairwise cluster probabilities $W = \{w_{rt}\}$ with $(r, t) = 1, 2, ..., k$. The corresponding entry in the adjacency matrix $A_{ij}$ is $1$ with probability $w_{rt}$ and $0$ otherwise. In the simplest version of the model (the planted SBM), all off-diagonal elements of the matrix $W$ are the same and equal to $w_{out}$, while all diagonal elements of $W$ are equal to $w_{in}$:
\be
W_{rt} =
\begin{cases}
w_{in}, \quad \text{$r=t$} \\
w_{out}, \quad \text{$r \ne t$}.
\end{cases}
\label{winout}
\ee
The assortative community structure corresponds to $w_{in}>w_{out}$. In the connectome context, the neurons belonging to the same cluster have a preferentially higher probability to be connected with a link than the neurons from different clusters. Still, some of the neurons within the same cluster in the structural connectome are not connected (clusters are not always cliques), allowing one to make use of stochastic models.

Importantly, for SBMs there is a certain threshold on the minimally allowed difference $\Delta w = w_{in} - w_{out}$ between the probabilities in order for the cluster structure to be resolved \cite{decelle2011asymptotic,decelle11}. Following conventional notation, let us introduce the rescaled cluster affinities, $c_{in}=Nw_{in}$ and $c_{out}=Nw_{out}$, which scale linearly with the number of the inner and outer edges of a typical community. The detectability rule suggests that the SBM clusters are asymptotically resolved ($N \gg 1$) as long as
\be
c_{in} - c_{out} > k \sqrt{c},
\label{thresh}
\ee
where $c=(c_{in}+c_{out})/2$ is the average of $c_{in}, c_{out}$. For dense networks $c_{in}, c_{out}, c \sim O(N)$ and, thus, condition \protect \eq{thresh} is satisfied at any small $\Delta w>0$. In the sparse case, $c \sim O(1)$, the threshold \protect \eq{thresh} provides a practically important condition on the parameters $\Delta w$ and $k$ for the cluster structure to be resolved.

Spectral methods, such as Laplacian, adjacency or modularity, have been widely used to uncover the community structure in relatively dense stochastic block model networks \cite{newman_sparse,newman2006modularity,newman06,nadakuditi12,reluct,Von2007tutorial}. The leading non-trivial eigenvectors of the corresponding operators provide dimensionality reduction of the system and these latent coordinates are then used by some conventional clustering algorithm (such as k-means) to perform partitioning into specified number of clusters \cite{Von2007tutorial}. However, as it was noted in \cite{Krzakala2013spectral}, for sparse networks the leading eigenvectors become uncorrelated with true community structure above the theoretical threshold \protect \eq{thresh}. This is because of the abundance of star-like sub-graphs (hubs) in a sparse network, which are identified by these operators instead of cyclic subgraphs associated with the internal structure of communities. Indeed, as these operators are related to random walks on a graph, true clusters interfere with hubs in their spectrum. As a result, it turns out that the spectral methods that exploit random-walk-related operators (such as modularity, adjacency or Laplacian) fail to find communities in rather sparse networks, despite of the network parameters satisfying the detectability condition \protect \eq{thresh}.

To overcome this difficulty, the spectrum of the Hashimoto matrix $B$ can be utilized, which is a transfer matrix of non-backtracking walks on a graph. It is defined on the edges of the directed graph, $i \to j, k \to l$, as follows
\be
B_{{i\rightarrow j},{k\rightarrow l}} = \begin{cases}
               A_{ij} A_{kl}\,\, \text{if}\,\, j=k\,\, \text{and}\,\, l\neq i\\
               0\,\, \text{otherwise},\\
            \end{cases}
\label{back}
\ee
It is seen from \eq{back} that the non-backtracking operator prohibits returns to the point which a walker visited at the previous step, thus effectively circumventing localization on the hubs. Notably, matrix $B$ is non-symmetric and has a complex spectrum. For Poissonian graphs, the spectrum of $B$ is constrained within a circle in the complex plane, whereas real eigenvalues of $B$ lying out of the circle are relevant to the community structure even in sparse networks. Associating the corresponding eigenvectors with the network partitioning allows detecting communities all the way down to the theoretical limit \protect \eq{thresh}. Interestingly, a "reluctant" version of the non-backtracking operator allows exploring the hanging trees of the network \cite{reluct}, which the original operator $B$ ignores by construction.

In \cite{newman_sparse} the corresponding flow operator was proposed, which conserves the probability flow at each step of the non-backtracking walker (see Fig. \ref{fig:pipeline}):
\begin{equation}
	F_{i \rightarrow j, k \rightarrow l} = \frac{\delta_{jk} (1 - \delta_{li})}{d_j - 1},
\label{flowmat}
\end{equation}
where $d_j$ is the degree of the vertex $j$. While the powers of non-backtracking matrix $B$ count the non-backtracking walks of particular length on a graph, the flow matrix $F$ is the transfer matrix of the non-backtracking probability. Similarly to the non-backtracking matrix, the bulk of the spectrum of $F$ lies in the complex plane within a circle of the radius
\begin{equation}
\label{eqn:critical_radius}
r=\sqrt{\frac{\langle d (d-1)^{-1}\rangle}{\langle d \rangle}},
\end{equation}
but, as shown in \cite{newman_sparse}, has a more clear edge of the spectral band. Importantly, the amount of isolated eigenvalues in the spectrum of the flow matrix corresponds to the number of clusters in SBM network \cite{newman_sparse}. In what follows, we will use the flow matrix \eq{flowmat} for the purpose of the connectome clustering.

The flow matrix $F$ defines the non-backtracking probability flow along the edges. While one is interested in the classification of the nodes, the eigenvectors of $F$ have to be carefully translated from the space of edges to the space of nodes. This is conventionally performed using the relation between the quadratic forms of modularity and flow operators \cite{newman_sparse,polovnikov20}.
From this correspondence one can see that contribution $u_i$ to the $i$-th node of the graph comes from the in-flow along all the directed edges adjacent to $i$. This procedure can be formally written as follows
\be
u_i = \sum_{j} A_{ij} v_{j \to i}^{F}
\label{rel}
\ee
where $v_{j \to i}^{F}$ is the component of the eigenvector of the flow matrix, corresponding to the directed edge $j\to i$. The element of the adjacency matrix $A_{ij}$ is non-zero as long as there is an edge between $i$ to $j$. Using \eq{rel} one can switch from edges to nodes representation of the non-backtracking flow and perform clustering of the nodes, e.g. using k-means on leading vectors $u_i$. Trivially, isolated vertices in a graph have undefined values of the flow, and they are not involved in graph clustering.

\section*{Clustering the connectome of a worm: how many clusters are detectable?}

The nervous system is one of the most complex parts of the nematode \textit{C.elegans} as the neurons constitute one third of all cells in this organism. The graph of the hermaphrodite connectome consists of $N=302$ vertices representing neurons and $C=4887$ edges (chemical synapses, see Methods) between them representing structural connections, as recorded in the new dataset \cite{cook2019whole}. Since only $11\%$ of the theoretically possible number of edges are present in the network, one may conclude that we deal with a rather sparse network. In the old dataset, in contrast, the network density is less than $5\%$, thus increasing the role of sparsity, as we will see below.

In order to obtain communities in \textit{C.elegans} connectome we implemented the spectral clustering approach, based on the leading non-trivial eigenvectors of the non-backtracking flow matrix  \protect \eq{flowmat}. The spectrum of the actual network corresponding to the new data is shown in the Fig. 1. Its isolated part, which is essential for the spectral clustering, consists of the maximal (trivial) eigenvalue and $9$ smaller eigenvalues that lie on the real line outside the bulk, constrained by a circle of the radius \protect \eq{eqn:critical_radius}. This picture suggests that there are $10$ communities in the network, encoded in the corresponding eigenvectors \cite{newman_sparse}. As we further found, the amount of isolated eigenvalues is invariant in both datasets analysed (see Fig. S2), despite the two-fold difference in the network density. This implies the existence of a stable mesoscopic structure, as revealed by the non-backtracking spectrum. 

\begin{figure}
    \centering
    \includegraphics[width=\columnwidth]{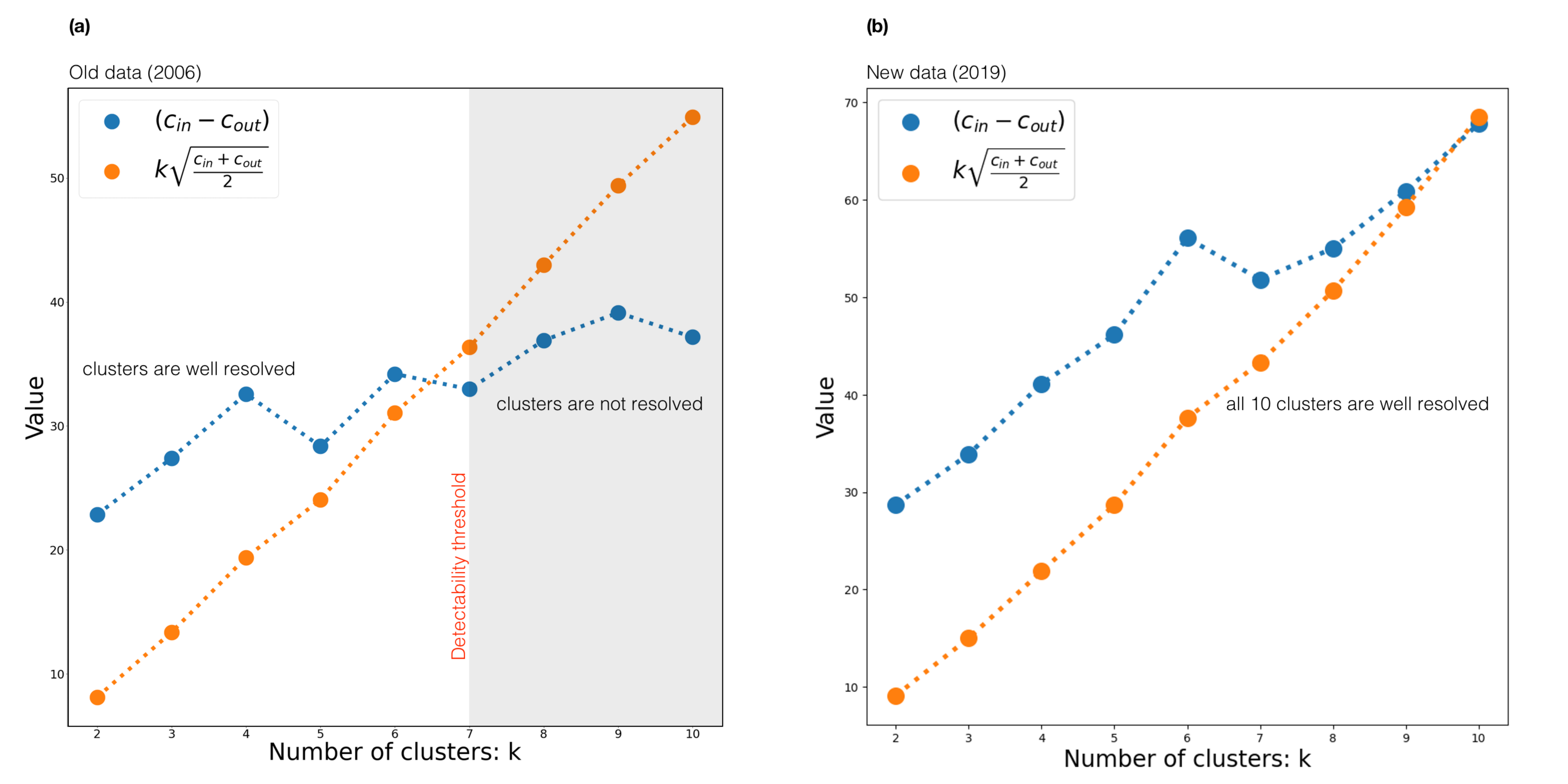}
    \caption{Graphical representation of the condition \protect \eq{thresh} as a criterion for the optimal number of clusters that can be detected (\textbf{a}) in the old  \cite{Chen2006wiring, varshney2011structural} and (\textbf{b}) new \cite{cook2019whole} connectome data. The intersection point of blue and orange curves provides the maximal amount of clusters $k_{max}$ at which the detectability condition \protect \eq{thresh} is satisfied.}
    \label{fig:clust_limit}
\end{figure}

Then we asked -- and this is not trivial -- how many clusters out of $10$ \textit{can be reliably resolved} in the given data. To this end, we suggest an approach based on the detectability threshold \protect \eq{thresh}. Namely, we note that for a given mean cluster strength $\Delta w$ the condition \protect \eq{thresh} establishes the maximum number of clusters that can be resolved in the sparse network of a given size $N$ and the average link probability $w=c/N$. Therefore, the critical number of clusters is related to the network parameters as follows
\be
k_{max} = \frac{\Delta w}{w} \sqrt{N}.
\ee
To find $k_{max}$ in the \textit{C.elegans} connectome we cluster the network into consecutive number of clusters $k = 2,3,4,..., 10$ using the eigenvectors of the non-backtracking flow matrix and determine $c_{in}$ and $c_{out}$ as the average amount of intra- and inter- links in the detected clusters. For each dataset we can determine the maximal amount of clusters $k_{max}$, such that $c_{in}-c_{out}$ is still greater than $k\sqrt{c}$ (see the sketch in Fig. \ref{fig:clust_limit}, explaining the procedure). Thus, the resulting value of $k_{max}$ provides the number of detectable communities, according to the detectability condition \eq{thresh}. An hierarchy of resulting community structures for different $k$ is shown in Fig. \ref{fig:pipeline}. 

While the total number of isolated eigenvalues is invariant in both datasets, the detectability condition clearly suggests a strong sensitivity of the amount of resolvable clusters $k_{max}$ to the network density of the sparse experimental data (see Fig. \ref{fig:clust_limit}). In the old and incomplete connectome data \cite{Chen2006wiring, varshney2011structural} only $k_{max}=7$ clusters can be resolved, thus, the remaining $3$ modules cannot be established due to strong sparsity of the data. At the same time, based on the same detectability condition applied to the new connectome data \cite{cook2019whole} with doubled density of synaptic connections, we conclude that all $k_{max}=10$ communities can be reliably recovered using the information from the flow matrix eigenvectors.

%Surprisingly, we found that the overlap between the obtained partitions for different $k$ and the anatomic partitions into ganglia has the maximum at $k=8$, Fig. 3c (green curve). We measured the agreement between various partitions by means of the adjusted mutual information (AMI, see Methods). Despite the ganglia were not expected to provide the absolute ground truth for the clusters of the structural connectome, the very fact that $k=7$ clusters provided the best agreement with the anatomic partition supports the proposed network-based inference approach and provides cross-validation to the optimal number of clusters that could be resolved in the \textit{C.elegans} connectome.

\subsection*{Non-backtracking flow outperforms other spectral methods in clustering of connectomes}

Having found the detectable number of modules, we next compared the performance of the non-backtracking flow matrix with traditional clustering operators, such as the normalized Laplacian and modularity matrix, on artificial networks with statistical properties similar to the experimental dataset \cite{cook2019whole}. To this end, we generated a family of stochastic block models with blocks similar to the ones we have obtained in the \textit{C.elegans} connectome. Namely, we fixed the network size, $N=279$, the outer-cluster probability, $w_{out}= 0.05$, and the total number of clusters, $k=7$. Furthermore, the sizes of the simulated blocks were chosen to match the sizes of the clusters in the original data. The only parameter subject to variation was the inner-cluster probability, $w_{in} = \{0.05, ..., 0.22, ..., 0.6\}$. For each value of $w_{in}$ we generated $200$ random SBMs. We ran the spectral clustering on $k=7$ leading eigenvectors of four operators: non-backtracking flow, normalized Laplacian, Laplacian and modularity.

The partitions predicted by the four operators were then assessed using the AMI scores, see Fig. S1. The results demonstrate distinctively better performance of the non-backtracking flow and normalized Laplacian over the modularity and Laplacian in prediction of the ground truth cluster structure of the simulated SBMs. The flow operator slightly outperformed the normalized Laplacian, especially in the region of intermediate relative cluster strengths, $w_{in}/w_{out} \approx 4-5$, which correspond to the empirical value. Such a moderate difference in prediction scores between the two operators is the result of a small network size $N$. As it was previously shown, in the limit of large (asymptotic) networks the non-backtracking operator outperforms the normalized Laplacian as well \cite{Krzakala2013spectral}. Notably, the empirical value $w_{in}/w_{out} \approx 4.4$, labeled by red dots on the AMI curves, is close to the detectability limit, as in this critical region AMI abruptly decays to zero (see Fig. S1). For large $N \gg 1$ there is an associated phase transition \cite{decelle2011asymptotic,decelle11}, thus highlighting criticality of the worm connectome. %In fact, this is a result of the maximal amount of clusters, chosen in accordance with the detectability condition \eq{thresh} above. Intermediate values of the prediction score reflect the balance between accurate annotation of the clustering structure and resolving the maximal possible amount of clusters.

\begin{figure}
    \centering
    \includegraphics[width=\columnwidth]{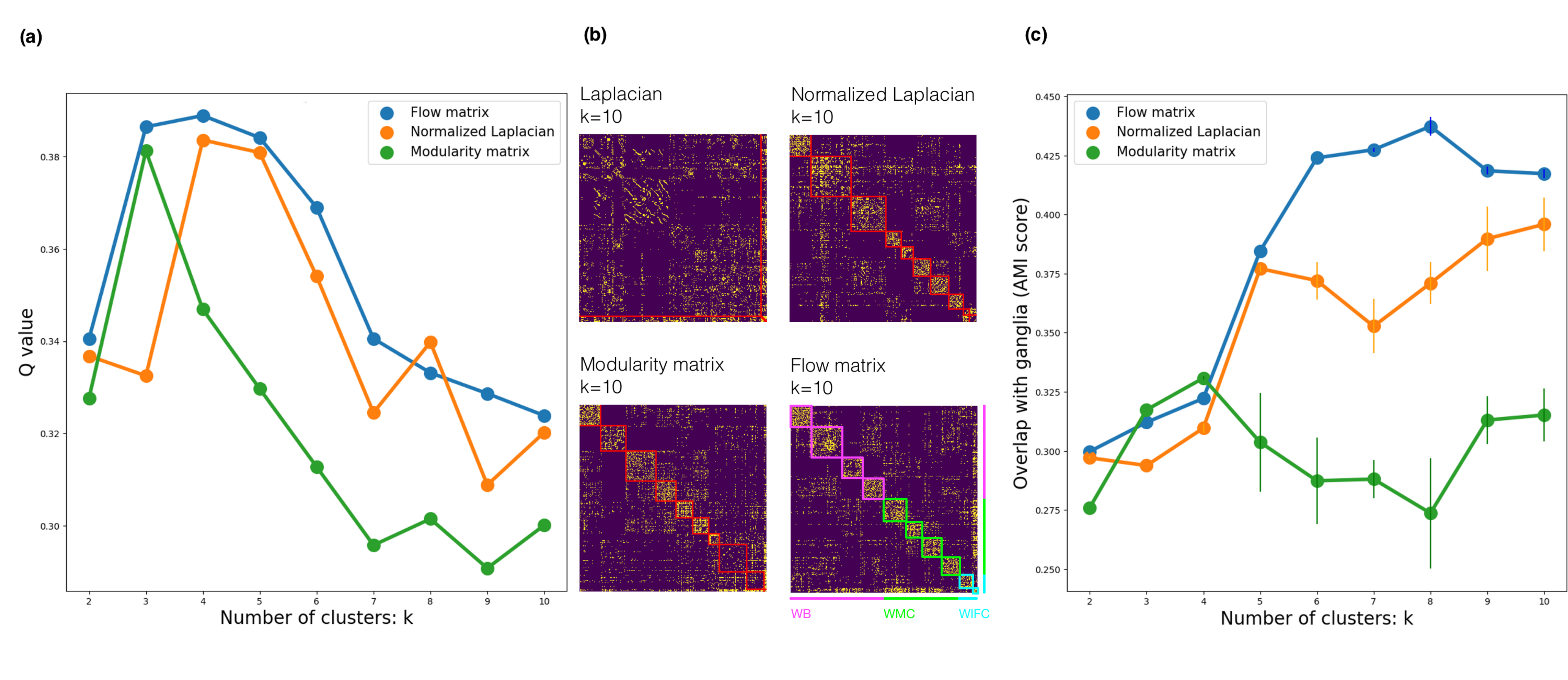}
    \caption{Quality comparison of different spectral approaches on the new connectome data \cite{cook2019whole}. (\textbf{a}) Q-values (\ref{mod1}) of partitions into $k$ clusters by various operators. (\textbf{b}) Partitions of the \textit{C.elegans} connectome into $k=10$ clusters obtained by different spectral methods. Each of the ten clusters inferred by the flow operator is annotated with a unique color according to its compartmental affiliation. (\textbf{c}) AMI score between ganglia and structural partitions for different number of clusters $k$.}
    \label{fig:SBM_AMI}
\end{figure}

Next we turned to performance analysis of different clustering algorithms on the experimental network. We compute the Q-scores for each of the annotations into clusters, which is equal to the modularity score of the partition and is defined on the basis of the Newman-Girvan modularity operator (see Methods). Using this metric we reveal a better quality of non-backtracking flow clusters for various values of the total number of clusters $k$ (see Fig. \ref{fig:SBM_AMI}a). As one can infer from Fig. \ref{fig:SBM_AMI}a and Fig. \ref{fig:SBM_AMI}b, for all $k$, except for $k=8$, the estimated quality of clustering by the flow operator outperforms other spectral approaches. In particular, the leading eigenvectors of the modularity matrix provide much poorer annotation into clusters, despite the quality metric being based on the modularity. This result is due to the sparsity issue discussed above; the leading spectrum of the non-backtracking flow matrix much better approximates the optimum of modularity function of a sparse graph than the leading spectrum of the modularity matrix itself. At the same time, we see that the normalized Laplacian produces annotations with similar, but steadily lower quality compared to the non-backtracking flow operator, in accord with the analysis of the SBM networks above (Fig. S1). Additionally, we have clustered the connectome using Infomap algorithm \cite{rosvall2009map}. Interestingly, the Infomap suggests that the optimal number of clusters equals to $k=3$ and provides a similar value of the modularity as obtained by the leading eigenvectors of the flow matrix ($Q \approx 0.38$), see Fig. 3(a). However, it fails to find all the clusters that evidently exist in the network (Fig. \ref{fig:clust_limit}).

To complement our analysis of Q-scores and better understand the (dis-)similarity of predictions by the flow matrix and other operators we further compute the pairwise relative overlaps between the clusters for the particular value of $k=10$ (see Fig. S3). With the overlap threshold in $80\%$ we find that 5 out of 10 modularity clusters poorly correspond to the flow matrix clusters, which translates into approximately $10\%$ difference in the Q-scores (Fig. \ref{fig:SBM_AMI}a). In the full agreement with the analysis above, only 3 clusters of normalized Laplacian overlap with the flow matrix clusters by less than $80\%$. Importantly, the smallest 10th cluster of the flow operator (consisting of as few as 7 neurons) is not resolved by normalized Laplacian. It is not fully resolved by modularity operator neither: despite an apparently high overlap with a modularity cluster (consisting of 4 neurons), the modularity fails to capture the other 3 neurons of the 10th flow matrix cluster. Indeed, tiny clusters in sparse networks present a particular challenge for traditional algorithms.

%Also, we have perform connectome clusterization using Infomap clustering algorithm \cite{rosvall2009map}. As a result, algorithm found $3$ clusters with the $Q = 0.38$.

\begin{figure}
    \centering
    \includegraphics[width=\columnwidth]{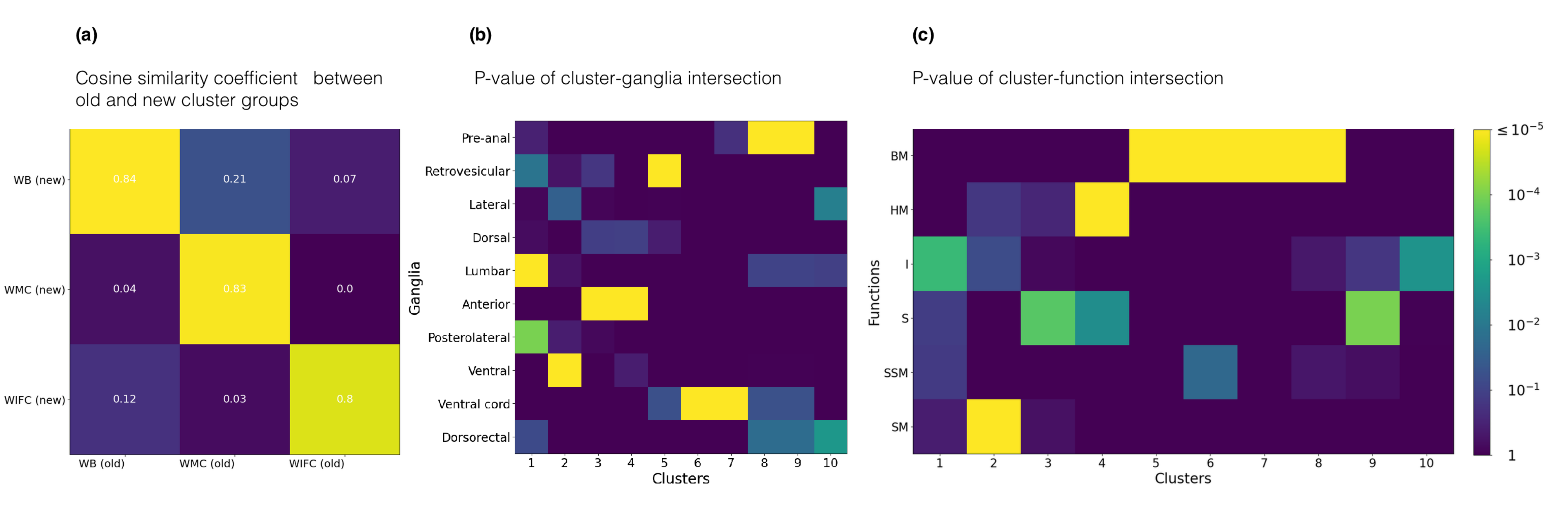}
    \caption{(\textbf{a}) Cosine similarity measure between WB, WMC and WIFC compartments for the old \cite{Chen2006wiring, varshney2011structural} and new connectome data \cite{cook2019whole}. (\textbf{b}) P-value of overlaps between the FM clusters and ganglia. 
    (\textbf{c}) P-value of overlaps between the FM clusters and functional groups (BM -- body motor neurons, HM -- head motor neurons, I -- interneurons, S -- sensory neurons, SSM -- sex specific motor neurons, SM -- sublateral motor neurons).}
    \label{fig:overlap_clusters_in}
\end{figure}

Additionally, we compare the structural clusters with ganglia. While ganglia represent the groups defined purely by anatomy and cannot be used as the ground truth for the structural partitions, they provide an important biological benchmark to estimate the technical noise produced by different clustering algorithms. We find that for sufficiently large number of clusters $k>4$, the AMI score for the non-backtracking flow takes the highest value compared to all other algorithms, Fig. \ref{fig:SBM_AMI}c. It is also worth noting that enrichment of the edges in the new dataset \cite{cook2019whole} has significantly increased the mutual information between the structural modules and the ganglia (Table S1). Still, similarity between the cluster groups obtained from old and new data is rather strong, see Fig. \ref{fig:overlap_clusters_in}a. Furthermore, as Fig. \ref{fig:overlap_clusters_in}b suggests, all flow matrix clusters in the new data have a statistically significant overlap with at least one ganglia ($p \leq 10^{-3}$).

As another biological benchmark, we consider partitioning of the \textit{C.elegans} nervous system into six groups corresponding to different neuronal functions:  motor neurons (head, body, sublateral, sex specific), sensory neurons, interneurons (see \href{https://docs.google.com/spreadsheets/d/1BfLIDWDIWga1uLQvBrxoqyU0OWrvBiCIWec1ptBKxyc/edit?usp=sharing}{Google Spreadsheet} and Fig. \ref{fig:overlap_clusters_in}c). Noticeably, three functional types (BM, HM and SM) are located within a particular group of clusters ($p \leq 10^{-5}$), as derived by the flow matrix; body motor neurons belong to 5th-8th clusters, head motor neurons mostly locate in the 4th cluster, sublateral motor neurons locate in the 2nd cluster. The group of interneurons belongs to the 1st and the 10th cluster (at $p \approx 10^{-3}$) and sensory neurons are spread between the 3rd, 4th and 9th clusters.

Put together, the statistical analysis of clusters obtained on real and simulated connectomes suggests that the non-backtracking flow operator outperforms conventional clustering approaches on networks of size and composition similar to the \textit{C.elegans} connectome.

%\section*{Neurons of different functional types correspond to particular structural modules}

%The above analysis demonstrates that all flow matrix clusters can be associated with either particular ganglia (except 4th) and/or dominant functional role. The 10th cluster is a special, since it is the smallest one (Fig. \ref{fig:SBM_AMI}b): it has only $7$ neurons and consists of almost all command interneurons of the worm (except AVB neurons pair) and can be relatively well explained by the lateral ganglion (it is the largest ganglion in the C.elegans connectome). It is also noticeable that the group of clusters from 5th to 8th, significantly overlapping with the first four ganglia, respond to the worm movements (body motor neurons). %The interplay between anatomy, functions and neuronal interconnections motivate

\section*{Comparison with previously reported connectome modules}

It is instructive to compare the results of the our algorithm (flow matrix) with other partitions reported in the literature. Here we analyse the results obtained by different approaches on the old dataset \cite{Chen2006wiring}, since to the best of our knowledge there have been no attempts to cluster the new connectome data in the literature.

Two open-source alternative annotations are considered, which were obtained by two different algorithms: iterative modularity maximization (IMMA) \cite{pan2010mesoscopic} and Erdos-Renyi mixture model (ERMM) \cite{pavlovic2014stochastic}. The IMMA approach is based on maximization of the modularity score ($k=6$ modules were found for the weighted connectome); the ERMM is a non-deterministic algorithm that fits an arbitrary, not planted, SBM network to the real connectome data ($k=9$ modules were found). First we compute the Q-scores of the clusters as obtained by the algorithms and compare to the quality of clusters produced by the flow matrix. We find that the flow matrix clusters produce a significantly higher quality of partition with the modularity value of $0.32$, which is to be compared with $0.27$ and $0.19$ for ERMM and IMMA clusters, respectively.

Next, we compute the mutual information between partitions predicted by various methods and biological benchmarks. 
As Table S1 demonstrates, the IMMA algorithm yields worse agreement with the ganglia (AMI$=0.31$) as compared to the flow matrix clusters (AMI$=0.34$). At the same time, the ERMM approach has the same AMI score with ganglia (AMI$=0.34$) and the mutual information between the ERMM and flow matrix clusters is sufficiently high (AMI$=0.45$). Also both the flow matrix and ERMM algorithm grouped command interneurons from the lateral ganglion together $(p \leq 10^{-5})$. Besides that, motor neurons were united into structural clusters by all three methods, see Fig. S5. All three methods successfully split parts of sensory neurons and interneurons, but only flow matrix was able to isolate the polymodal neurons $(p \leq 10^{-4})$. 

Overall, we conclude that the previous partitions of \textit{C.elegans} connectome, based on the old dataset \cite{Chen2006wiring}, display lower quality of clustering (Q-values) and worse biological interpretability than can be achieved using the leading part of the spectrum of the non-backtracking flow matrix. 

%Importantly, all three algorithms produce structural partitions that are more similar to each other than to the anatomic ganglia. Thus, despite structural partitions are notably different from the anatomic one, we conclude that the flow matrix and ERMM better agree with the biological benchmark than IMMA.

%A more detailed analysis between particular ganglia and clusters in the old data reveals that the 1st, 6th, and 7th ganglia do not overlap significantly with any structural cluster in all the three clustering approaches (flow matrix, ERMM and IMMA) (see Fig. S5a). Though it it tempting to refer such an agreement between various methods to some biological peculiarities of posterolateral, dorsal and retrovesicular ganglia, we see that in the new connectome data all ganglia overlap with at least one flow matrix module (Fig. \ref{fig:overlap_clusters_in}b). Accordingly, the overall AMI score between ganglia and modules in the new data is evidently higher (Table S1).

\section*{Cross-talk between the clusters is determined by neuronal programs}

\begin{figure}
    \centering
    \includegraphics[width=\columnwidth]{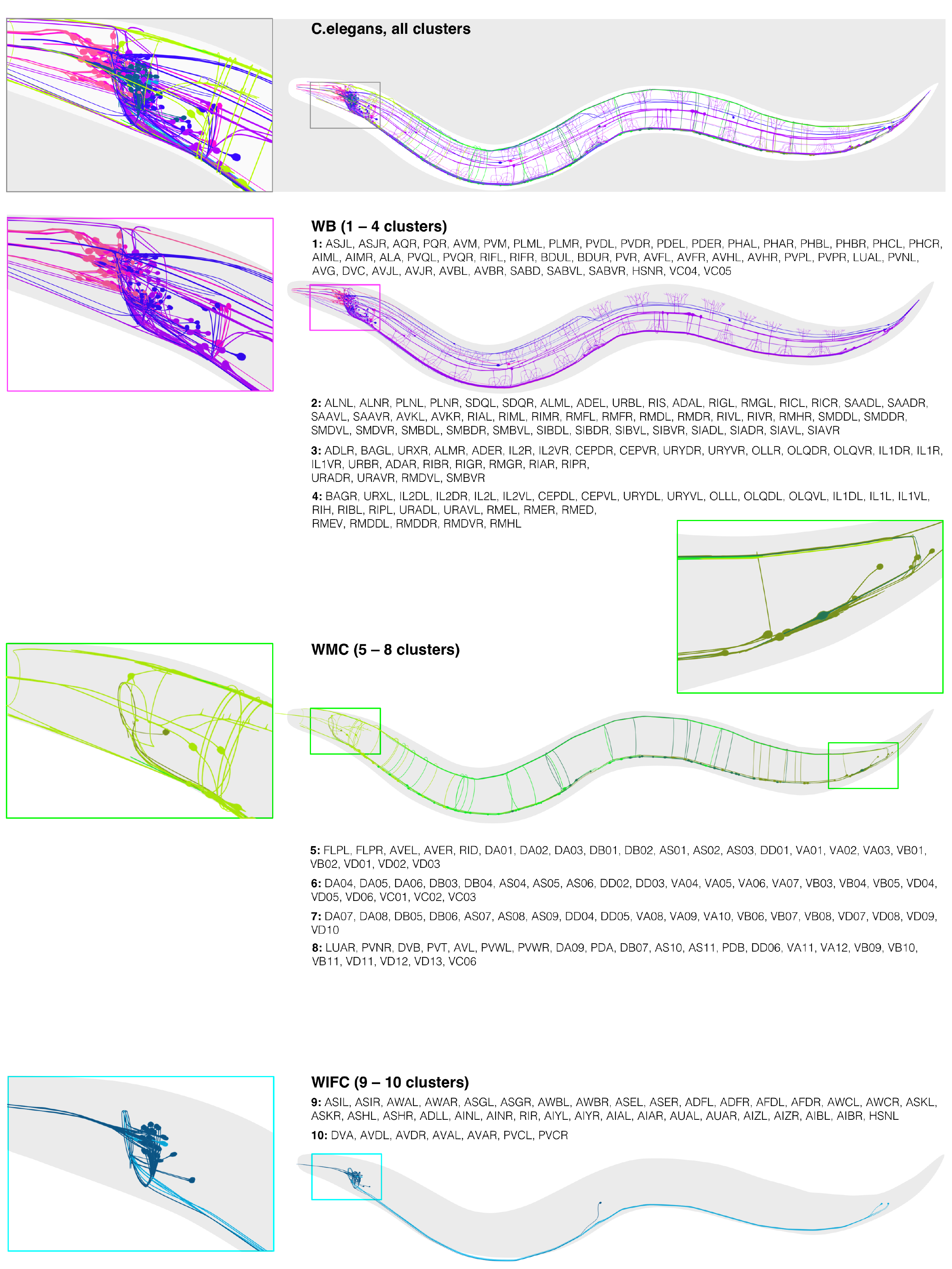}
    \caption{Three-dimensional visualization of the detected clusters in the connectome. On the basis of the complementary analysis of functional roles and anatomical locations, we identify three compartments: WB, WMC, WIFC. Original $3$D coordinates were taken from \href{http://canopus.caltech.edu/virtualworm/}{the Caltech Wormbase project} \cite{harris2010wormbase}.}
    \label{fig:bio_int}
\end{figure}

On the basis of the complementary nature of detected communities (functional roles and anatomical locations), we reveal the following neuronal \textit{compartments}. 
This classification is supported by similar neuronal functions and/or 3D coordinates in the worm body: the Worm Brain (1st-4th); the Worm Movement Controller (5th-8th); the Worm Information Flow Connector (9th and 10th). It should be noted that the neurons listed below are the names of the neuron sets, for example, VA contains twelve individual neurons VA1-VA12 or ADA contains ADAL and ADAR. For detailed description of cluster elements see Fig. \ref{fig:bio_int}. 

\begin{itemize}

\item Worm Brain (1st, 2nd, 3rd and 4th clusters)

Similarly to the multifunctional organization of the brain of more complex organisms, we found that neurons in these four clusters had a common anatomical position and were involved in complex multimodal processes \cite{white1986structure}. Based on the corresponding 3D model (see Fig \ref{fig:bio_int}), one can note that the 3rd and 4th clusters are closer to the nose of the worm and the 1st and 2nd clusters are located behind them. This anatomical similarity between the clusters is consistent with their close functions. Accordingly, all polymodal neurons of the head part of the worm (IL1, IL1D, IL1V, OLQD, OLQV, RIM, ASH) and sensory neurons of the anterior ganglion (IL2D, IL2, IL2V, OLL) are located in the 3rd cluster. Wherein bilaterally symmetric neurons split between these two clusters, thus $27$ out of $28$ neurons of the 3rd cluster are right neurons (R), while $21$ out of $29$ neurons of the 4th cluster are left ones (L) (see \href{https://docs.google.com/spreadsheets/d/1BfLIDWDIWga1uLQvBrxoqyU0OWrvBiCIWec1ptBKxyc/edit?usp=sharing}{Google Spreadsheet}). 

The majority of interneurons (ADA, RIA, RIB, RIF, RIG, RIH, RIR, RIS, RIV, URB, URX, SAAD, SAAV, AVK, SDQ, SIAD, SIAV, SIBD, SIBV) belong to the 1st and 2nd clusters. Such a "layered" organization is consistent with intuition about how the signals received by the worm's nervous system should be processed. 

%From the perspective of the compressed connectome view (Fig. \ref{fig:graph_model}), clearly seen, that most of the contacts from the 1st cluster ($\approx 80\%$) connect it with the 4th cluster which in turn has high contact probability ($\approx 16\%$) with the smallest and full of command neurons 7th cluster. This fact could mean that the pair of 1st and 4th clusters plays the roles of acceptor and preprocessor of the sensory information flow with subsequent signal transmission to command neurons in 7th cluster in order to further correction of motor programs, through the dense contacts of the 7th cluster with in 2nd and 6th clusters.

\item Worm Movements Controller (5th, 6th, 7th and 8th clusters)

These four clusters contain the ventral cord neural group, which is split between them according to the anatomical positions of the neurons. Namely, neurons in the head half fall into the 5th and 6th clusters, while neurons in the tail body half fall into the 7th and 8th clusters (Fig. \ref{fig:bio_int}). 

Together, almost $83\%$ of neurons in these four clusters belong to the ventral cord (AS, DA, DB, DD, VA, VB, VC, VD, AVE), which motoneurons are located along the entire body of the worm and split exactly into two groups in accordance with which half of the body of the worm these neurons belong to: 5th and 6th clusters correspond to the head part and 7th and 8th clusters are responsible for the tail movements (bright and dark green colors Fig. \ref{fig:bio_int}). 

The remaining $17\%$ are represented by neurons of the tail ganglia: pre-anal ganglion, and dorsorectal ganglion (PV, LUA, PWV, PDA, PDB, DVA, DVB, DBC) and belong to the 8th cluster, which is responsible for the tail part of the ventral cord (Fig. \ref{fig:overlap_clusters_in}b). Excitatory motor neurons in the ventral cord function as motor rhythm generators and underlie body undulation during reversal and forward movements \cite{wen2018caenorhabditis}. That is why we refer this pair as a \textit{worm movements controller}. The connection probabilities between these four clusters are reasonably low ($ \approx  5\%$ in average), which is tenable if we interpret them as disjoint parts of the movement control system, and the dense connections between motor neurons from opposite parts of the worm body are not functionally significant. %At the same time, a strong connectivity between these four clusters and the clusters 1st, 2nd, 9th and 10th full of interneurons logical consistent and equal ($ \approx 30\%$ in average), where 10th cluster consists of command neurons and that agrees with the notion that command neurons are responsible for coordinating the movements of the worm. %(Fig. \ref{fig:graph_model}). 

\item Worm Information Flow Connector (9th and 10th clusters)

Almost $53\%$ of the neurons in these two clusters belong to the lateral ganglion: sensory neurons (ADF, ADL, ASE, ASG, ASH, ASI, AFD, AWA, AWB, AWC, ASJ, ASK), interneurons (AIA, AIB, AIN, AIY, AIZ, AUA, AVJ) and command interneurons (AVA, AVD, PVC). Command interneurons, which are located in the 10th cluster, by definition receive a convergence of integrative sensory inputs and output to a multifarious group of pattern-generating efferent neurons \cite{kawano2011imbalancing}. This is consistent with the contact probabilities between the 10th cluster and the Worm Movements Controller. For example, there is an evidence that ablation of AVB or AVA command neurons leads to impairment of spontaneous forward or backward movements \cite{wicks1996dynamic}, suggesting they are one of the most critical regulators for the directional motion. 

The distribution of contacts between the 10th command cluster and other clusters clearly shows its significant role in the information flow integration processes: it receives information about the outer environment from the interneurons located in the 1st, 2nd and 9th clusters (Fig.  \ref{fig:overlap_clusters_in}c) and coordinates the behavior of the worm through dense contacts with the worm movements controller. Therefore, one can propose that the 10th cluster plays the role of a "command post" coordinating movement of a worm (clusters 5th-8th), and responsible for the implementation of the worm's motor programs. 

%Also we note that the cluster 10 forms a topological rich club and there are only AVB and AVE pairs of command neurons missing from it (they are lying in the 1st and the 5th clusters correspondingly). The rich club command neurons play a role of connector hubs in the C.elegans connectome, with high betweenness centrality, and many intermodular connections to neurons in different modules \cite{towlson2013rich}.}

%In terms of sensory neurons and interneurons of the lateral ganglion, located in the 9th cluster, it should be noted that the vital role of the lateral ganglion consists in the processing of sensory information and providing an essential connection between the sensory and motor components of the \textit{C.elegans} nervous system \cite{chatterjee2007understanding}. 

\end{itemize}

%The three groups of clusters have been identified in both old and new connectomes and their pairwise comparison demonstrates significant similarity of the found groups, that means remarkable clusterization result invariance with respect to network density (cosine similarity between the respective cluster groups is $>0.8$, Fig. \ref{fig:overlap_clusters_in}(a)). While poor connectivity information from the old connectomic data set wasn't enough to exactly uncover the underlying network community structure and find all possible clusters, additional accounting information about synaptic contacts provided in \cite{cook2019whole} allowed to allocate 3 additional clusters.}

\section*{The non-backtracking connectome clusters largely correspond to the contactome modules}

In two recent studies \cite{brittin2021multi, moyle2021structural} a network of $10^5$ membrane contacts (the contactome) from the \textit{C.elegans} nerve ring was generated and analysed. The contactome data contains information about spatial neurite contacts, thus providing an important biological benchmark to compare the modules of the structural connectome.

\begin{figure}
    \centering
    \includegraphics[width=\columnwidth]{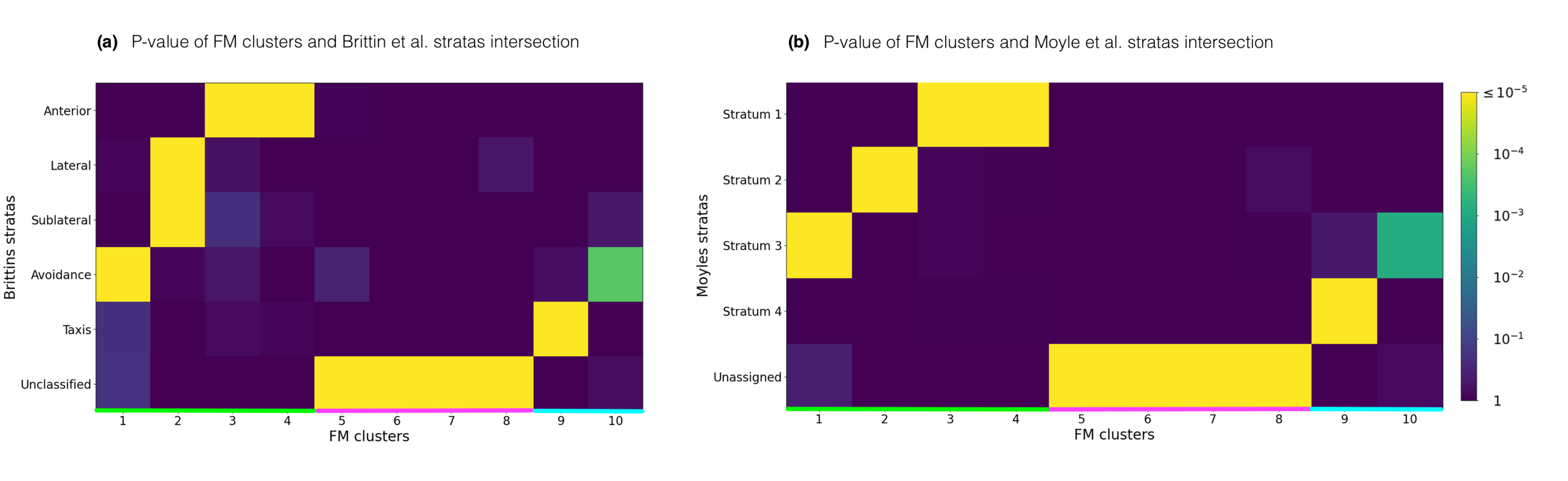}
        \caption{P-values of the overlaps between the non-backtracking flow clusters and contactome modules: (a) Brittin et al. stratas \cite{brittin2021multi} or (b) Moyle et al. stratas \cite{moyle2021structural}.}
    \label{fig:contacome_p_value}
\end{figure}

The authors in \cite{brittin2021multi, moyle2021structural} have clustered the contactome using classical methods into several neuronal groups or \textit{strata} that we statistically compare with the flow matrix clusters (see Fig. \ref{fig:contacome_p_value}a,b). We find that all clusters of the Worm Brain compartment (1-4) have a significant intersection with four contactome stratas (Anterior, Lateral, Sublateral, Avoidance) \cite{brittin2021multi}; the p-value of the intersection is $(p \leq 10^{-5})$. The same analysis for the second contactome dataset from \cite{moyle2021structural} shows a similar result: the Worm Brain clusters intersect significantly with the first three strata. 

Interestingly, the Taxis strata \cite{brittin2021multi} and Stratum 4 \cite{moyle2021structural} have a statistically significant overlap with our 9th cluster $(p \leq 10^{-5})$; our 10th cluster has a statistically significant overlap with Avoidance strata \cite{brittin2021multi} and Strata 4 \cite{moyle2021structural} $(p \leq 10^{-4})$. These are the clusters comprising the Worm Information Flow Connector with vast majority of the neurons belonging to the head of the worm. Since only head neurons comprised the contactome dataset (nerve ring), the other neurons of the worm were left unassigned there. Accordingly, we see that the clusters 5-8 from the Worm Movements Controller (WMC) have significant intersections with Unclassified and Unassigned strata from the contactome datasets. All together, we conclude that the contactome modules significantly overlap with the connectome clusters of brain neurons that fall into the Worm Brain and Worm Information Flow Connector.

Next, we asked whether the same quantitative correspondence between contactome and connectome modules can be established using other clustering algorithms. For that we calculate the AMI score between the corresponding partitions, see Fig. S4 and Table S1. The maximal value of AMI score is achieved for 6 clusters, since the neurons from the other 4 clusters are missing in the contactome dataset. Indeed, as Fig. 6 shows, the clusters from 5 to 8 contain neurons that are left unassigned in both datasets of the contactome. Accordingly, at $k=6$ the flow matrix yields connectome clusters that demonstrate the best quantitative agreement with the contactome modules among all the clustering methods (flow matrix: AMI = $0.51$ for the Brittin et al. and AMI = $0.47$ for the Moyle et al.; IMMA algorithm: AMI = $0.4$ for the Brittin et al. and AMI = $0.43$ for the Moyle et al.; ERMM algorithm: AMI = $0.41$ for the Brittin et al. and AMI = $0.36$ for the Moyle et al.). Furthermore, as Fig. S4 suggests, the flow matrix clusters appear to correspond better to the contactome modules than the normalized Laplacian and modularity clusters (e.g., the FM clusters overlap by more than 20\% stronger with the contactome modules than the clusters produced by the other two spectral methods). 

This analysis highlights that the non-backtracking algorithm yields interpretable connectome communities that strongly overlap with the contactome modules derived from the neurite spatial contacts. The obtained non-backtracking communities correspond to the neurite contacts much better than the previously reported clusters of \textit{C.elegans} connectome.

%We point out that this is an important manifestation of the sparsity of the connectome network, which is not taken into account by the classical spectral algorithms. Thus, this analysis highlights that the non-backtracking algorithm provides a considerably better correspondence with the contactome modules compared to other clustering methods due to the intrinsic sparsity associated with the connectome network.

\section*{Conclusion}

In this paper we performed a detailed analysis and demonstrated applicability of the spectrum of non-backtracking random walks to the problem of the structural connectome clustering on the model example of \textit{C.elegans}. This clusterization method outperforms traditional clustering methods on simulated models in the regime of low density of connections, virtually circumventing the sparsity issue and related emergence of hubs. Our analysis of the \textit{C.elegans} connectome has shown that already on a relatively small network size ($N=279$) the communities produced by the non-backtracking method -- as compared to previously reported modules in the connectome -- result in a better clustering quality and biological interpretability in the terms of functions and neurite spatial contacts.  

We applied the non-backtracking clustering scheme for two versions of the \textit{C.elegans} connectome (old \cite{Chen2006wiring} and new \cite{cook2019whole}). While in both datasets $k=10$ isolated eigenvalues can be seen in the spectrum of the non-backtracking flow matrix, the detectability condition suggests that only in the new connectome all $k=10$ clusters can be reliably detected (see Fig.\ref{fig:clust_limit}). Based on the complete \textit{C.elegans} connectome \cite{cook2019whole}, we reveal ten interpretable communities that we further classify into three distinct compartments: (i) four multifunctional head clusters full of ring neurons ('Worm Brain'); (ii) four clusters responsible for movements control in the head and tail halves of the worm ('Worm Movements Controller'); (iii) one cluster made up from the command interneurons and one cluster from the lateral ganglion consisting of sensory neurons and interneurons ('Worm Information Flow Connector'). 

%In contrast to the previously reported partitions in the literature, the non-backtracking operator revealed that \textit{all} ganglia have statistically significant overlap with the structural modules. Namely, in addition to previously reported ganglia, ventral and dorsorectal anatomic regions were shown to exhibit a significant overlap with the structural modules. 

%This result importantly suggests that the anatomically defined regions \textit{mediate} propagation of information through the interconnected modules along the worm body. 

Comparison with the recently mapped contactome modules reveals strong interpretability of the non-backtracking communities in terms of the neurite spatial contacts. We find that the 'Worm Brain' compartment (clusters 1-4) consists of the neurons with axons that project into the anterior part of the nerve ring, while the 'Worm Information Flow Connector' (clusters 9-10) compartment stores the amphid neurons that project axons into the posterior part of the worm head. Since the other 'Worm Movements Controller' compartment (clusters 5-8) contains the neurons that are unclassified in the contactome, we conclude that contactome modules strongly reproduce the connectome clusters. Such a result supports Peter's rule as a principle of synaptic specificity in the \textit{C.elegans} brain.

%In a recent study by Cook et al. \cite{cook2023neuronal} it was proposed that if neurons reach a threshold of physical adjacency (an edge in the contactome network), they are likely to form a synaptic connection (an edge in the connectome adjacency matrix). 

%In more detail, algorithmic detection of command neurons, especially considering that they form a topological rich club \cite{towlson2013rich}, clearly shows that the flow matrix based decomposition has been able to capture both modular and core-periphery aspects \cite{borgatti2000models} of the \textit{C.elegans} connectome mesoscale organization. 

%The resulting system of modules has a distribution of connections among themselves, which is consistent with the assumed organization of information flows inside the nervous systems of the worm \cite{white1986structure} (Fig. \ref{fig:graph_model}).

Broadly, our study highlights deep interconnections between anatomical locations (metric space embedding), mesoscopic structure (topological embedding) and biological functions of the neurons that determine behaviour of an organism. In order to derive these relationships accurately it is important to precisely resolve the topological modules in the connectome network. In the framework of the one of the simplest model organisms we demonstrate that the cluster analysis of the connectome should be performed by taking into account the intrinsic sparsity of the network, which is addressed by the non-backtracking operator informed by the corresponding detectability thresholds. Given universality of our approach, we believe it can be further extended to connectomes of more complex organisms with larger networks, where the effect of sparsity would be even more dramatic.

\section*{Methods}

\subsection*{Data}
We have worked with old and new open-access data from the \textit{C.elegans} connectome analysis project \cite{Chen2006wiring, varshney2011structural} and Cook SJ et al. paper \cite{cook2019whole}. Both datasets are updated and revised versions of the wiring data originally published in \cite{white1986structure}. Neuron interactions, locations, sensory endings, and neuromuscular junctions, as well as the structure of the connectome, have been well studied and have been found to be invariant with respect to the type of animal \cite{white1986structure, hall1991posterior}, however, there is now growing concern that the C.elegans connectome is not invariant \cite{yemini2021neuropal, brittin2021multi, witvliet2021connectomes}. Connectome 3D model used for the reconstruction of cluster elements anatomical positions was taken from \href{http://canopus.caltech.edu/virtualworm/}{the Caltech Wormbase project} \cite{harris2010wormbase}. 

The two versions of the connectome (\cite{Chen2006wiring, varshney2011structural} and \cite{cook2019whole}) have significantly different number of edges: in the refined data of Cook SJ \textit{et al} \cite{cook2019whole} there is almost twice more synaptic contacts as compared to the BL Chen \textit{et al} \cite{Chen2006wiring} previous work ($6334$ vs $2990$). The new hermaphrodite connectome from \cite{cook2019whole} is a network with $302$ vertices and $6334$ edges: $1447$ edges are formed by gap junctions only; $4887$ contain only chemical synapses. The old hermaphrodite connectome \cite{Chen2006wiring, varshney2011structural} is a network with $302$ vertices and $2990$ edges: $796$ edges are formed by gap junctions and $1962$ contain only chemical synapses.

The entire nervous system is broken down into two large disconnected components and two isolated neurons (CANL, CANR) and additionally the VC06 neuron is isolated in the old connectome data \cite{Chen2006wiring, varshney2011structural}. Twenty of the neurons in one of the components are located within the worm pharynx, which has its own separate nervous system, and the remaining $280$ (or $279$ for \cite{Chen2006wiring, varshney2011structural}) neurons (excluding two isolated neurons) are located in various ganglia along the worm body. During the preprocessing stage, all connections in the connectome are made undirected and unweighted. Furthermore, we have divided the graph into two subgraphs according to contact types: chemical synapses or gap junctions and analyzed the connectome formed only by the synaptic contacts, because these two types of connections are fundamentally different in nature and their functions are also distinct. 

There is a large body of knowledge on individual neurons that produce node-wise features. In this work, we have used the classification of neurons into ten anatomically defined ganglia (posterolateral, ventral, pre-anal, lateral, dorsorectal, dorsal, retrovesicular, ventral cord, anterior and lumbar ganglia) and six functional groups (body motor neurons, head motor neurons, interneurons, sensory neurons, sex specific motor neurons, sublateral motor neurons) from \cite{altun2005handbook, cook2019whole}.

As another benchmark for testing biological validity of our clusters, we have used the contactome adjacency matrices \cite{brittin2021multi, moyle2021structural}, because contactome itself contains information about axon position and metric structure of the C.elegans nervous system. Contactomes have fewer number of described neurons ($170$ neurons in the \cite{brittin2021multi} stratas and $181$ in the \cite{moyle2021structural}), therefore we added all missing neurons to the stratas \textit{Unassigned} and \textit{Unclassified} correspondingly. Full contactomes descriptions could be found on \href{https://docs.google.com/spreadsheets/d/1BfLIDWDIWga1uLQvBrxoqyU0OWrvBiCIWec1ptBKxyc/edit?usp=sharing}{Google Spreadsheet}.

%Generation of the SBM matrix family played a crucial role in the algorithm performance test. For it we used two probabilities: $w_{in}$ for the elements contained in the same cluster and $w_{out}$ for the elements lying in different communuties. The generated matrix family consisted of $4000$ matrices: $20$ different pairs of parameters $w_{in}, w_{out}$ and $200$ generations for each of them. The $6$th pair of $w_{in}, w_{out}$ parameters was taken from the clusterization of the C.elegans connectome by flow matrix, more specific from the partition of the seven clusters. For each of the $20\times 200$ matrices different clusterization algorithms (Flow matrix, normalized Laplacian, Modularity) have been used and their performance compared by \textit{adjusted mutual information} score (see mean score results Fig. \ref{fig:SBM_AMI}).

\subsection*{From stochastic block model to non-backtracking random walks}

One of the most popular methods for community detection (in particular, of the connectome \cite{pan2010mesoscopic}) is optimization of modularity. In fact, it can be shown that the generalized modularity functional provides the entropy of a Poisson weighted stochastic block model with quenched degrees (configuration model). Such models, for example, describe the results of single-cell contact counting experiments in chromatin networks, as was shown by us recently \cite{polovnikov20}. If the degrees of all vertices $d_i=\sum_j A_{ij}$ are kept fixed, without additional imposed cluster structure, the expected weight of the edge under random degree-preserving randomization is simply $P_{ij}=\frac{d_i d_j}{\sum_{i} d_i}$ for $i \ne j$. Assuming that the stochastic blocks are superimposed over the configuration model, each entry $A_{ij}$ of the adjacency matrix of the observed network becomes a Poisson random variable with the mean $P_{ij} w_{rt}$, such that the nodes $i$ and $j$ are assigned to the groups $G_r$ and $G_t$, respectively. Thus, the total statistical weight of $A$ conditioned on the cluster probability matrix $W$, quenched degrees $d_i$ and group labels $g_i$ can be factorized into the product of the Poisson probabilities and written down as follows
\be
{\cal{Z}} (A | W, d_i, g_i) = \prod_{i<j} \frac{P_{ij} w_{g_i g_j}^{A_{ij}}}{A_{ij}!} \exp\left(-P_{ij} w_{g_i g_j}\right)
\ee
which produces the following entropy
\be
S_{conf.} \propto \log {\cal{Z}} (A | W, d_i, g_i) = \sum_{i<j} \left(A_{ij} - \gamma P_{ij}\right) \delta_{g_i g_j}
\label{entr}
\ee
where $\gamma$ is some parameter that depends on $w_{in}$ and $w_{out}$ of the planted SBM \eq{winout} as follows
\be
\gamma = \frac{w_{in}-w_{out}}{\log w_{in} - \log w_{out}}
\label{gamma}
\ee
Clearly, the entropic functional \eq{entr} up to the parameter $\gamma$ is nothing but the \textit{modularity functional}, which is widely used in clustering tasks, for connectome clustering as well \cite{pan2010mesoscopic}. It is important to note that generally the parameter $\gamma$ have to be chosen self-consistently with the cluster parameters of the partition \eq{gamma}, for which the iterative procedure has been recently proposed \cite{polovnikov20}.

Modularity optimization has been originally proposed and proved to be useful for clusterization of scale-free networks, since, as we have shown above, it explicitly conserves the scale-free property of the degree distribution under stochastic randomization. Although most of the real-world networks are scale-free, modularity is one of the most popular approaches in spectral clustering. However, if one relaxes the degrees preservation assumption, the background probability becomes uniform $P_{ij}=p$ and the underlying graph is assumed to be simply a $G(N, p)$ Erdos-Renyi graph. Then the second term in \eq{entr} does not depend on cluster labels of the nodes, and maximization of the entropy for a given amount of clusters corresponds to maximization of the adjacency functional
\be
S_{ER} \propto \log {\cal{Z}} (A | W, g_i) = \sum_{i<j} A_{ij} \delta_{g_i g_j}
\label{adj}
\ee
which is trying to maximize the internal weight of the clusters. In a more general problem setting of a manifold learning, one is looking for the optimal representation (embedding) of $N$ vertices in a low-dimensional space described by a set of coordinates $g_i, i=1,2,..., N$ (suppose, the latent space is one-dimensional for simplicity). As long as close points in the original high-dimensional space should be eventually put close in the latent space, the natural functional to be minimized is
\be
S_{ML} \propto \log {\cal{Z}} (A | W, g_i) = \frac{1}{2}\sum_{i\ne j} A_{ij} \left(g_i-g_j\right)^2
\ee
which can be written as a quadratic form of the graph Laplacian, $L=D-A$
\be
S_{ML} \propto \sum_{i,j} L_{ij} g_i g_j
\label{lap}
\ee
Of course, a similar functional over latent coordinates can be written for the modularity functional \eq{entr} as well.

Thus, we see that statistical inference of the optimal cluster structure is associated with optimization of a certain functional over partition of graph nodes. However, finding the global maximum of \eq{entr},\eq{adj},\eq{lap} is a very difficult computational task. To overcome this difficulty, spectral methods are used, which rely on the fact that the most essential information about the optimal partition is encoded in the first non-trivial eigenvectors of the corresponding operator. Indeed, the quadratic form associated with the manifold learning problem can be approximated by projecting the coordinates to the leading eigenvectors of the operator.

\subsection*{Modularity matrix}

In \cite{newman2004finding} the modularity matrix of a graph was defined as

\begin{equation}
    M := A - \frac{d d^T}{2C},
\end{equation}

where $A$ is the adjacency matrix, $d = (d_1, \dots, d_n)^T$ is the \textit{degree-vector} comprised of the vertices degrees and $C = \frac{1}{2}\sum_{i=1}^n d_i$ is the total number of edges in the network. 

We computed a quantitative measure of modularity for each partition of graphs into several communities, using the standard Newman's modularity (Q value):
\begin{equation}
	Q := \frac{1}{2C} \sum_{i,j}\left(A_{i,j} - \frac{d_i d_j}{\sum_i d_i}\right) \delta_{g_i g_j}
\label{mod1}
\end{equation}
By notation, $A$ is the adjacency matrix of connectome ($A_{ij} = 1$, if neurons $i, j$ are connected, and $0$, otherwise). The degree of each vertex $i$ is given by $d_i= \sum_{j} A_{ij}$. $C$ is the total number of edges on the connectome graph, equal to $C=\frac{1}{2}\sum_i d_i$ and $\delta$ is the Kronecker delta and $g_i$ is the label of the community to which vertex $i$ is assigned. As we see, \eq{mod1} is different from the entropic functional \eq{entr} by a particular normalization coefficient used.

\subsection*{Laplacian and normalized Laplacian}

Laplacian is widely used in spectral manifold learning methods, a framework known as Laplacian Eigenmaps. The graph Laplacian matrix is defined as
\begin{equation}
    L := D - A,
\end{equation}
where $A$ is the adjacency and $D$ is the diagonal matrix of degrees. Though Laplacian is related to many physical phenomena, such as heat propagation, a more direct connection with random walks is provided by the Normalized Laplacian (or Random Walks Laplacian), $L_{RW}=D^{-1}L$, which is also frequently used for clustering. Note that $L_{RW}$ is non-symmetric, however, its spectrum is real. Obviously, $L_{RW}$ has the same set of eigenvalues as the symmetric normalized Laplacian
\begin{equation}
    L_{norm} := D^{1/2} L_{RW} D^{-1/2}=D^{-1/2} L D^{-1/2} = I - D^{-1/2} A D^{-1/2}.
\end{equation}

%More information on standard spectral clustering methods can be found in this article \cite{Von2007tutorial}.

\subsection*{Similarity measures}

In order to assess the similarity between different partitions and biological benchmarks we use the adjusted mutual information score (AMI), defined as follows. Suppose that we have a set $S$ and two partitions of $S$: $U$ and $V$, the elements of the partitions are called clusters. Let us denote the probability that some random object falls into a cluster $U_{i}$ of $U$ as $P_{U(i)}$ which is equal to $\frac{|U_i|}{|S|}$. The entropy calculated for the partition $U$ is equal to $H(U)=-\sum_{i=1}^R P_U(i)\log P_U(i)$. Using the introduced notation, we can express the mutual information for $U$ and $V$ as
\begin{equation}
    MI(U,V) :=\sum_{i=1}^R \sum_{j=1}^C P_{UV}(i,j)\log \frac{P_{UV}(i,j)}{P_U(i)P_V(j)}.
\end{equation}
Importantly, this measure of similarity tends to be larger when the two partitions have a larger number of clusters even when we use the same number of elements for clustering. To avoid such biases one can use the adjusted mutual information which is defined as
\begin{equation}
     AMI(U,V) := \frac{MI(U,V)-E\{MI(U,V)\}}{\max{\{H(U),H(V)\}}-E\{MI(U,V)\}},
\end{equation}
where $E\{MI(U,V)\}$ is the expected value of the mutual information of $V$ and $U$.

Therefore, AMI is 0 when the similarity is equal to its expected value under random permutation of the vertices between the groups and 1 for identical partitions.

\typeout{}
\bibliographystyle{unsrt}
\bibliography{references}

\section*{Acknowledgements}
All authors thank A. Gorsky and E. Burnaev for stimulating discussions on the project and A. Moiseeva for help with illustrations. This work was supported by the Russian Science Foundation, Grant №21-75-30024. AO, AC and KP acknowledge the support of Idea Foundation grant in collection and analysis of the connectome data

\section*{Additional information}
The authors declare no competing interests.

\section*{Author contribution}
K.P. conceptualized the study and design. Data collection and analysis was performed by A.O. and A.C. All authors participated in writing

\section*{Code availability}
Source code and data available on \href{https://github.com/ArseniiOnuchin/worm_clusterization}{GitHub project page}. 

\section*{Data availability}
Additional table with $280$ C.elegans neurons and their functions distributed among flow matrix clusters (obtained from old \cite{Chen2006wiring, varshney2011structural} and new \cite{cook2019whole} connectome data) is available on \href{https://docs.google.com/spreadsheets/d/1BfLIDWDIWga1uLQvBrxoqyU0OWrvBiCIWec1ptBKxyc/edit?usp=sharing}{Google Spreadsheet}.

%\title{Non-backtracking walks reveal functional communities in sparse structural connectome of a worm}
\title{Supplementary Information \\ Communities in C.elegans connectome through the prism of non-backtracking walks}

\author{Arsenii A. Onuchin, Alina V. Chernizova, Mikhail A. Lebedev, Kirill E. Polovnikov}

\maketitle

\makeatletter
\renewcommand \thesection{S\@arabic\c@section}
\renewcommand\thetable{S\@arabic\c@table}
\renewcommand \thefigure{S\@arabic\c@figure}
\makeatother
\setcounter{figure}{0}

In this document, we provide 5 Supplementary Figures and 1 Table that additionally illustrate the results of the clustering analysis of the \textit{C.elegans} structural connectome from the main text. In the first two figures (Figs. S1, S2) we discuss the technical performance of the flow matrix algorithm of clusterization and the properties of its eigenvalues spectrum. In the next figure (Fig. S3) we provide pairwise intersections between the clusters, found by three spectral methods: Normalized Laplacian, Modularity matrix and Flow matrix. In the following two figures (Figs. S4, S5) we provide an additional biological justification to the found clusters.  The supplementary table contains information about the pairwise overlaps between various structural clusters and biological modules. All the data used in the paper is deposited on this link: \href{https://docs.google.com/spreadsheets/d/1BfLIDWDIWga1uLQvBrxoqyU0OWrvBiCIWec1ptBKxyc/edit?usp=sharing}{Google Spreadsheet}.

\begin{figure}
    \centering
    \includegraphics[width=300pt]{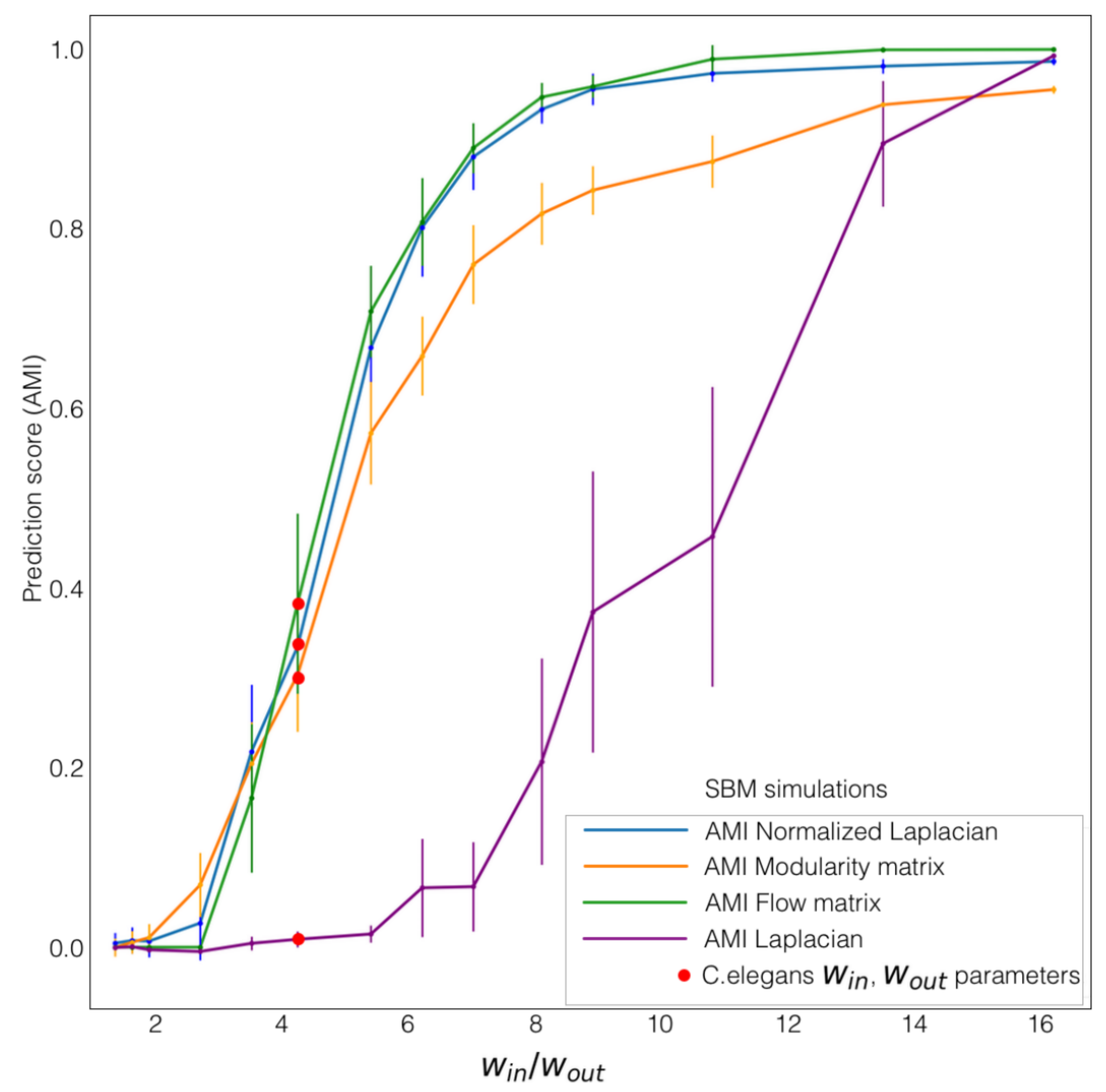}
    \caption{Mean Adjusted Mutual Information (AMI) score for simulations with the planted Stochastic Block Model (SBM) with weight parameters $w_{in}$, $w_{out}$, size $N=279$ and number of clusters $k=7$. The AMI score is averaged over $200$ realizations of random SBM graphs with fixed parameters (the error bars reflect the corresponding statistical error). For each realization of a random graph AMI score is computed between the underlying SBM partition (the ground truth) and clusters inferred in that graph by four different operators: Laplacian, normalized Laplacian, modularity operator and flow matrix. The red dots denote the empirical value $w_{in}/w_{out} = 0.22/0.05 \approx 4.4$, corresponding to the connectome (data from Chen et al, 2006 [34,35]). }
    \label{fig:SBM_AMI_sep}
\end{figure}

\begin{figure}
    \centering
    \includegraphics[width=\columnwidth]{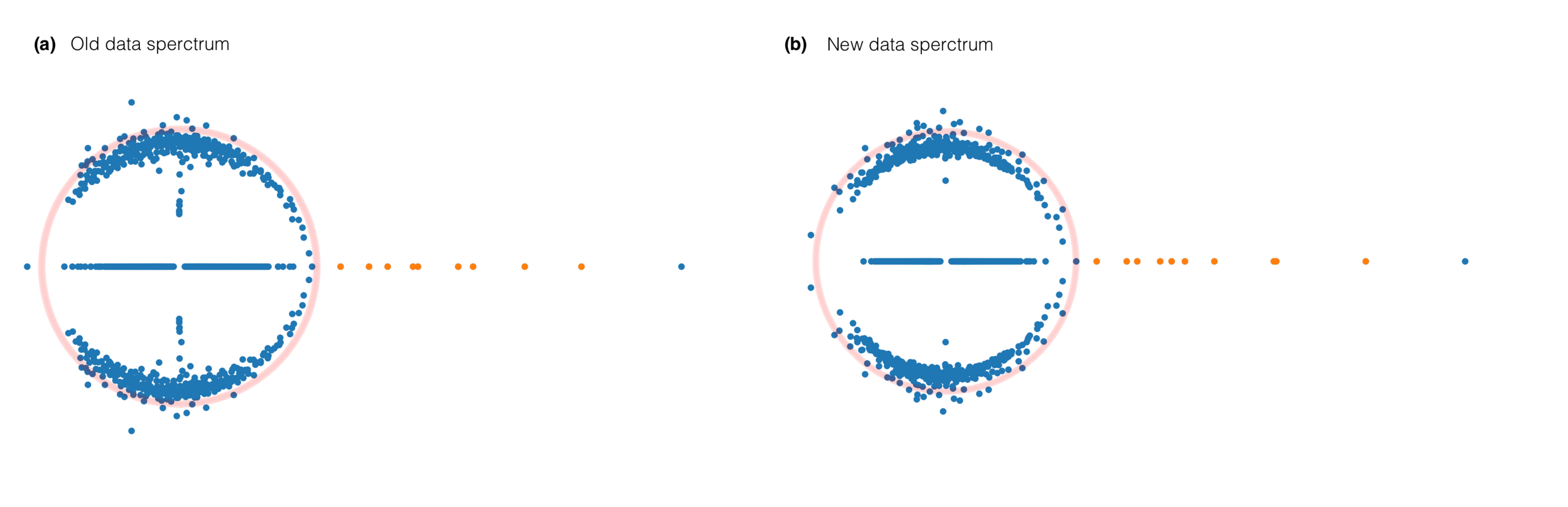}
    \caption{Comparison of the eigenvalues spectra of the flow matrix computed for two datasets: \textbf{(a)} Chen et al, 2006 [34, 35] ("old data") and \textbf{(b)} Cook et al, 2019 [25] ("new data "). In both cases the spectrum consists of complex eigenvalues constrained within a circle of radius $r$ (see Eq. (5) of the main text) and a set of isolated eigenvalues on the real axis (orange) together with the leading eigenvalue (blue). Despite the 2-fold increase of the total number of edges in the new dataset, the amount of isolated eigenvalues remains the same, $k=10$.}
    \label{fig:old_new_spec}
\end{figure}

\begin{figure}
    \centering
    \includegraphics[width=\columnwidth]{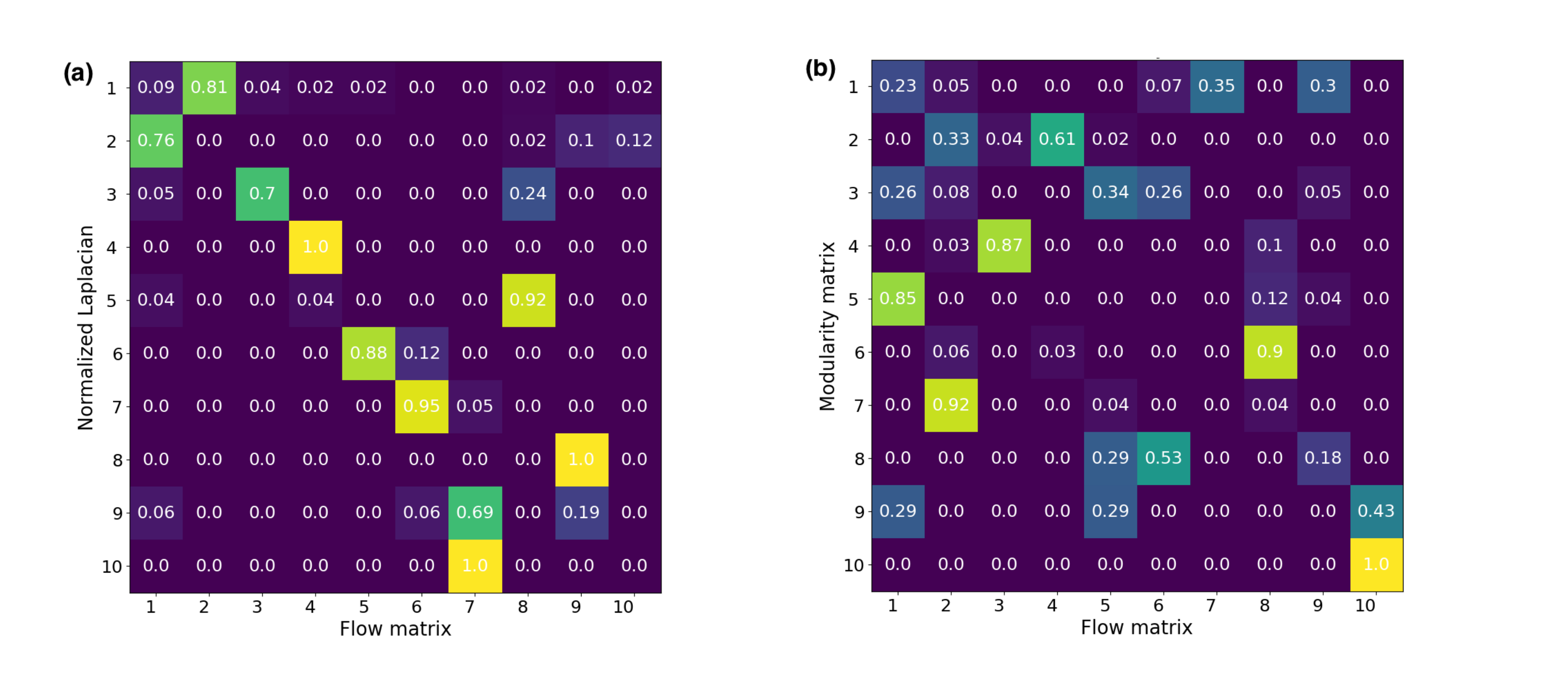}
    \caption{This figure shows the percent of the cluster, which was obtained by Normalized Laplacian or Modularity matrix spectral methods, contained in the corresponding Flow matrix cluster. (\textbf{a}) Overlaps between Normalized Laplacian and Flow matrix $10$ clusters  found in the new connectome data [25]
    (\textbf{b}) Overlaps between Modularity matrix and Flow matrix $10$ clusters  found in the new connectome data [25].}
    \label{fig:overlap_10_clusters_new_data}
\end{figure}

\begin{figure}
    \centering
    \includegraphics[width=\columnwidth]{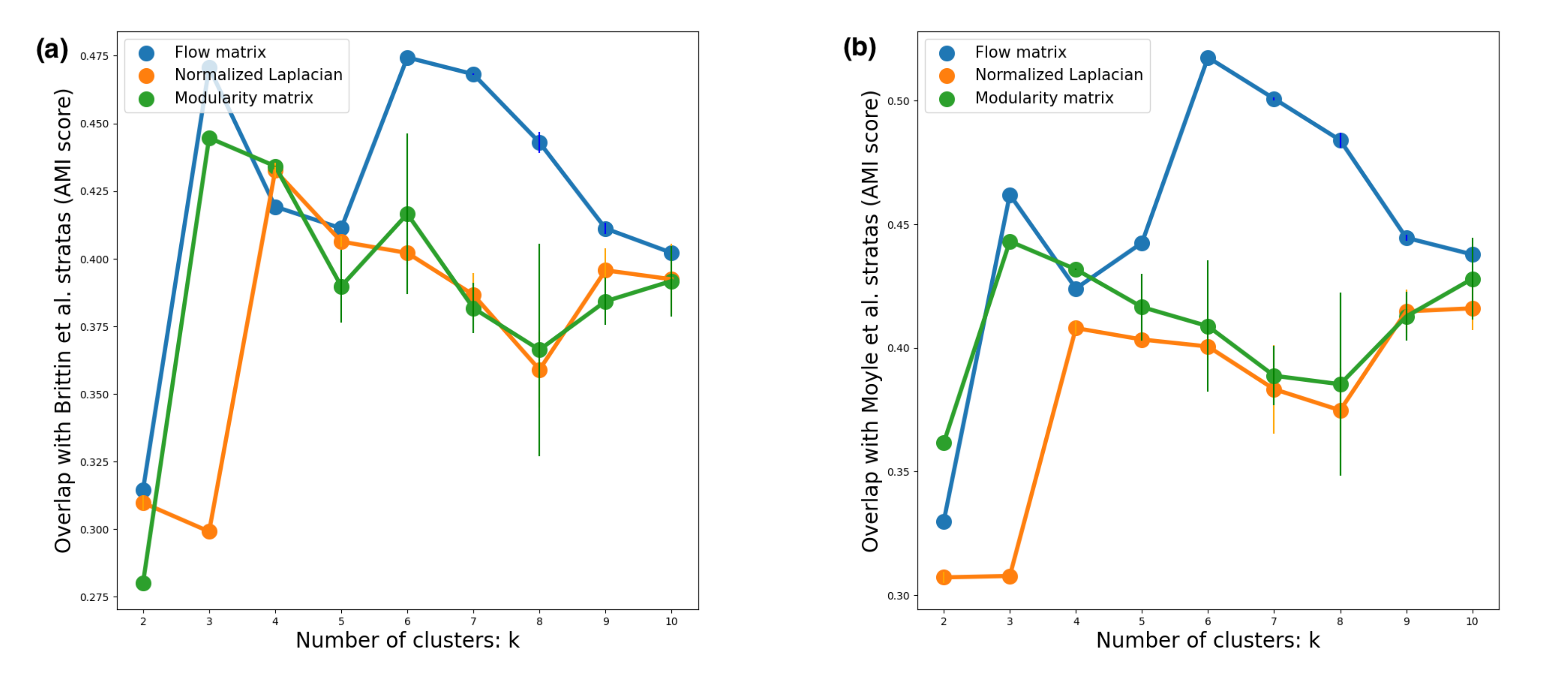}
    \caption{(a) The AMI score computed between the modules from Brittin et al. [29] and partitions of the structural connectome (the new data, Cook et al, 2019 [25]) into $k$ clusters, as inferred by three operators: flow matrix, normalized Laplacian and modularity matrix. The peak of AMI score for the flow matrix corresponds to $k=6$ modules identified in Brittin et al. [29]. (b) The same as in (a) but for $k=6$ modules from Moyle et al. [30].}
    \label{fig:contacome_AMI_scores}
\end{figure}

\begin{table}
\begin{center}
\begin{tabular}{ |l||l|  }
 \hline
 & AMI score\\
 \hline
 IMMA clusters [41] vs Ganglia (old data [34, 35]) &  $0.31$ \\
 \hline
 IMMA clusters [41] vs Moyle et al. [30]  (old data [34, 35])  &  $0.43$ \\
\hline
IMMA clusters [41] vs Brittin et al. [29]  (old data [34, 35])  &  $0.40$ \\
\hline
\hline
 ERMM clusters [33] vs Ganglia (old data [34, 35])  &  $0.34$ \\
 \hline
 ERMM clusters [33] vs Moyle et al. [30]  (old data [34, 35])  &  $0.36$ \\
 \hline
ERMM clusters [33] vs Brittin et al. [29]  (old data [34, 35])  &  $0.41$ \\
\hline
\hline
IMMA clusters [41] vs ERMM clusters [33]  (old data [34, 35])  &  $0.41$ \\
\hline
\hline
 Flow Matrix clusters (maximum for $8$ clusters) vs Ganglia   (new data [25])   & $0.425$ \\
 \hline
 Flow Matrix clusters vs Ganglia (old data [34, 35]) & $0.34$ \\
 \hline
Flow Matrix clusters (maximum for $6$ clusters) vs Moyle et al. [30] (new data [25]) &  \textbf{$0.51$} \\
\hline
Flow Matrix clusters (maximum for $6$ clusters) vs Brittin et al. [29] (new data [25]) &  \textbf{$0.47$} \\
 \hline
 Flow Matrix clusters vs IMMA clusters [32] (old data [34, 35]) &  $0.4$ \\
 \hline
 Flow Matrix clusters vs ERMM clusters [33]  (old data [34, 35]) &  $0.45$ \\
 \hline
 
\end{tabular}
\caption{The overlap (AMI score) values between modules found by various algorithms in the old [34, 35] and new [25] C.elegans connectomes (as indicated) and biological benchmarks (ganglia, contactome modules [29, 30]).}
\label{tab:diff_mod_struct}
\end{center}
\end{table}

\begin{figure}
    \centering
    \includegraphics[width=\columnwidth]{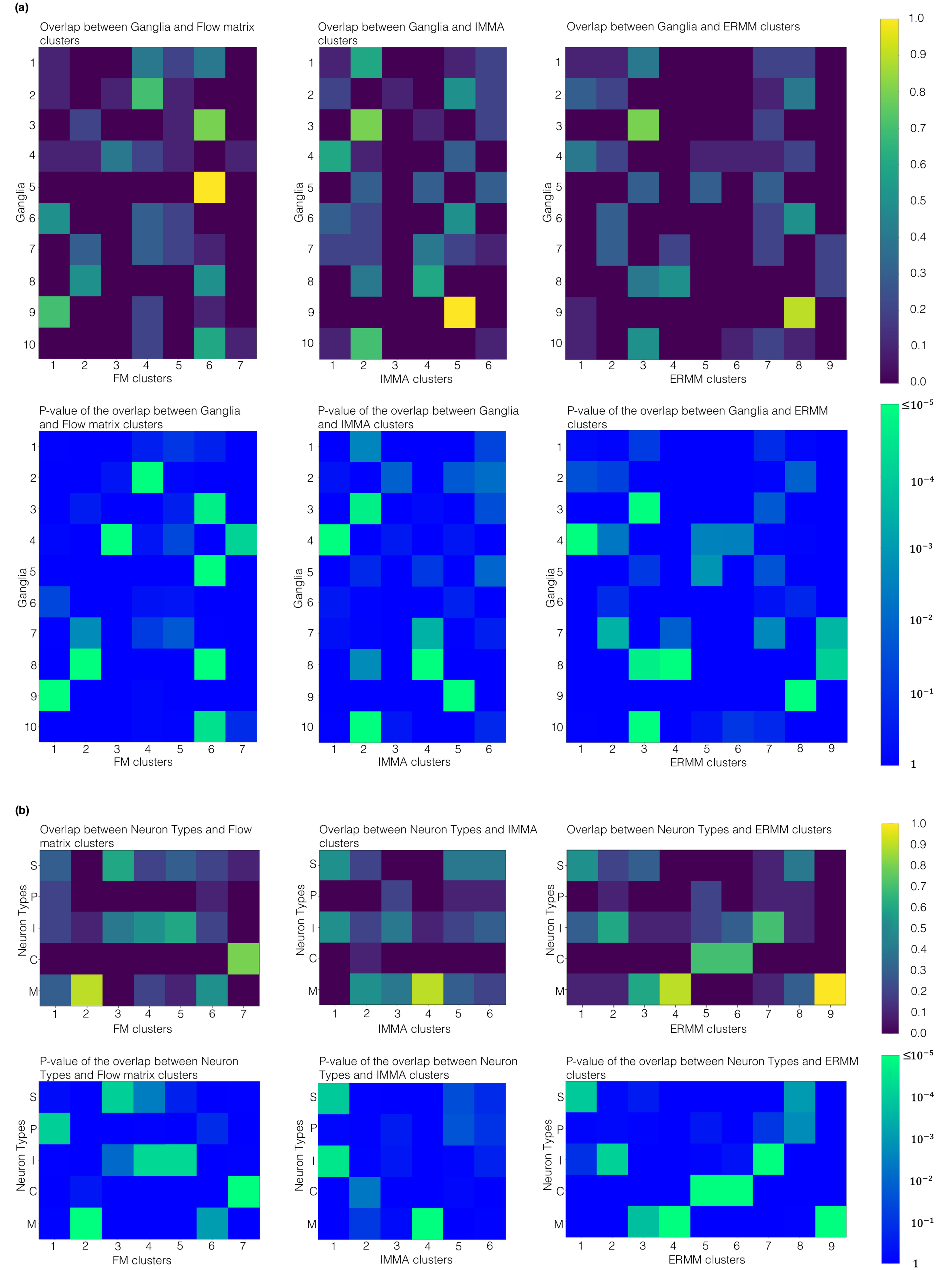}
    \caption{(\textbf{a}) Overlaps and p-values between ganglia and clusters found by three different algorithms: flow matrix (FM), iterative modularity maximization algorithm (IMMA) [41] and Erdos-Renyi Mixture Model (ERMM) [33]. The structural clusters are obtained on the old connectome data [34, 35]. 
    (\textbf{b}) Overlaps and p-values between the clusters (same as in (a)) and neuronal types: sensory neurons (S), polymodal neurons (P), interneurons (I), command neurons (C), motoneurons (M). }
    \label{fig:overlap_clusters_in_old}
\end{figure}

\end{document}